\documentclass[journal=ancac3,manuscript=article]{achemso}

\usepackage[version=3]{mhchem} 
\usepackage[utf8]{inputenc}
\usepackage[T1]{fontenc}
\usepackage{bm}
\usepackage{comment}
\mciteErrorOnUnknownfalse
\author{Kilian D. Stenning}
\affiliation{Blackett Laboratory, Imperial College London, London SW7 2AZ, United Kingdom}
\email{k.stenning18@imperial.ac.uk}
\author{Jack C. Gartside}
\affiliation{Blackett Laboratory, Imperial College London, London SW7 2AZ, United Kingdom}
\author{Troy Dion}
\affiliation{Blackett Laboratory, Imperial College London, London SW7 2AZ, United Kingdom}
\alsoaffiliation{London Centre for Nanotechnology, University College London, London WC1H 0AH, United Kingdom}
\author{Alexander Vanstone}
\affiliation{Blackett Laboratory, Imperial College London, London SW7 2AZ, United Kingdom}
\author{Daan M. Arroo}
\affiliation{London Centre for Nanotechnology, University College London, London WC1H 0AH, United Kingdom}
\author{Will R. Branford}
\affiliation{Blackett Laboratory, Imperial College London, London SW7 2AZ, United Kingdom}

\title  {Magnonic Bending, Phase Shifting and Interferometry in a 2D Reconfigurable Nanodisk Crystal}

\begin{document}


\section*{Abstract}

Strongly-interacting nanomagnetic systems are pivotal across next-generation technologies including reconfigurable magnonics and neuromorphic computation. Controlling magnetisation state and local coupling between neighbouring nanoelements allows vast reconfigurable functionality and a host of associated functionalities. However, existing designs typically suffer from an inability to tailor inter-element coupling post-fabrication and nanoelements restricted to a pair of Ising-like magnetisation states.

Here, we propose a class of reconfigurable magnonic crystal incorporating nanodisks as the functional element. Magnetic nanodisks are crucially bistable in macrospin and vortex states, allowing inter-element coupling to be selectively activated (macrospin) or deactivated (vortex). Through microstate engineering, we leverage the distinct coupling behaviours and magnonic band structures of bistable nanodisks to achieve reprogrammable magnonic waveguiding, bending, gating and phase-shifting across a 2D network. The potential of nanodisk-based magnonics for wave-based computation is demonstrated \textit{via} an all-magnon interferometer exhibiting XNOR logic functionality. Local microstate control is achieved here \textit{via} topological magnetic writing using a magnetic force microscope tip.

\section*{Keywords}
reconfigurable magnonic crystal, magnonics, nanomagnetism, metamaterials, artificial spin system, tunable coupling, microstate control \\

\noindent
Magnetic nanostructures form the basis of diverse next-generation device architectures, with applications including magnonic logic \cite{khitun2010magnonic,khitun2011non,chumak2015magnon}, non-Boolean computation \cite{romera2018vowel,torrejon2017neuromorphic,bhanja2016non,mizrahi2018neural,grollier2020neuromorphic,sangwan2020neuromorphic} and reconfigurable magnonic crystals (RMCs)\cite{grundler2015reconfigurable,haldar2016reconfigurable, krawczyk2014review, wang2017voltage, topp2010making, neusser2009magnonics, kruglyak2010magnonics, chumak2017magnonic,vedmedenko20202020,gliga2020dynamics} -  periodic arrays of magnetic nanoelements exhibiting tunable magnon (spin-wave) dynamics through varying magnetic configurations (microstates). Tailoring the nanoelement shape as well as the geometry of the underlying lattice offers a platform for tuning collective magnon dynamics to store and process information\cite{locatelli2014spin, lenk2011building}.

Reconfigurable control of information carried in spin-wave amplitude, frequency and phase is key for magnonic computation. Existing designs have delivered such control for waveguides\cite{pirro2014spin,garcia2015narrow}, multiplexers\cite{vogt2014realization,heussner2018frequency, davies2015field}, phase shifters \cite{hansen2009dual,ustinov2014nonlinear,kostylev2007resonant} and logic devices \cite{khitun2010magnonic,fetisov1999microwave,schneider2008realization,klingler2015spin,ding2012realization,hertel2004domain}, yet they typically require biasing \textit{via} current or magnetic fields which dissipate heat and may disturb adjacent magnetic states. In RMCs, biasing is provided by the magnetic shape anisotropy of the nanopatterned array elements. Different microstates allow different functionality such as band-gap creation\cite{krawczyk2014review}, amplitude and frequency modulation \cite{ding2012realization,haldar2016reconfigurable,dion2019tunable,iacocca2020tailoring,sadovnikov2018spin,wang2015tunable,semenova2013spin,ma2011micromagnetic,kim2009gigahertz}, magnon path bending \cite{haldar2016reconfigurable} or defect-induced phase shifts \cite{louis2016bias}. However, in existing designs microstate access is severely restricted – typically offering a handful of microstates from a vast 2\textsuperscript{N} macrospin state space in an $N$-element system. Furthermore, the number of available states increases dramatically when considering the curling of magnetisation at the nanoelement edges, further limiting microstate control. This presents a hard bottleneck to realising the power and potential of RMCs. Additionally, bending spin-waves in a 2D network whilst conserving magnonic information is a challenge in realising compact magnonic circuitry, with existing designs showing power loss and shifts in magnon frequency and phase\cite{vogt2012spin,davies2015towards,sadovnikov2018spin}.

Recent advancements in non-invasive selective nanomagnetic writing have used localised stray fields of magnetic tips to allow reprogammable definition of active magnon channels \cite{albisetti2018nanoscale} and writing of nanoelement magnetisation states \cite{wang2016rewritable,gartside2016novel,gartside2018realization} in addition to a proposed scheme achieving writing \textit{via} current-driven domain walls\cite{gartside2020current}, allowing complete microstate access in RMCs. Existing RMCs and writing schemes have revolved around nanowire based designs, constrained to single-domain macrospin states aligned along a shape-defined easy axis, hard-coded at the fabrication stage. The Ising-like nanowires restrict the available magnetic states and do not provide the capability to effectively switch off inter-element coupling by writing a flux-closure state.

An alternative to nanowires are circular-shaped nanomagnets (nanodisks), capable of supporting both macrospin and vortex magnetisation states\cite{cowburn1999single,jubert2004analytical,chung2010phase,metlov2008map,ostman2014hysteresis,gelvez2019coercive} [Fig. \ref{Fig0}], allowing for vast microstate engineering potential. In thin nanodisks, macrospin states comprise spins with a net alignment along a single macrospin axis, resulting in a net magnetisation and large dipolar field --- promoting strong inter-element coupling. Vortex states comprise a circulating in-plane magnetisation and central out-of-plane polar magnetisation, giving low dipolar-field leakage due to the circulating flux-closure\cite{chung2010phase} and hence weak inter-element coupling. Chirality and polarity of in-plane and out-of-plane regions may be set independently, allowing four vortex permutations. The difference in inter-element coupling between macrospin and vortex states allows definition of disks as ‘active’ --- participating in magnon transmission (macrospin), and ‘inactive’ (vortex), significantly enhancing freedom and functionality relative to nanowire designs.  
Existing methods of nanodisk state control involve global and pulsed\cite{van2006magnetic,yamada2007electrical} magnetic fields, introducing asymmetry \cite{uhlivr2013dynamic,agramunt2014controlling,dumas2011chirality} and coupling to neighbouring nanomagnets \cite{haldar2015vortex}. These methods add additional complexities to device fabrication, break disk symmetry  and do not allow full control of any arbitrary disk in an array. Controlling disk states will advance their  potential for key technologies including neuromorphic computation \cite{romera2018vowel} and RMCs\cite{PhysRevLett.107.127204,huber2011ferromagnetic,kaffash2020control,louis2016bias,kumar2014magnetic,shibata2003dynamics,sugimoto2011dynamics,vogel2011coupled,jung2011tunable,barman2016enhanced,mondal2020magnetic,mondal2020spin,burgos2019time,ramasubramanian2020tunable,morales2020ultradense,li2020second} and support development of schemes\cite{bhanja2016non} leveraging their bistable magnetisation states.

To address these challenges, through micromagnetic simulation we propose a class of RMC incorporating nanodisks as the functional element.  \textit{Via} the rotational freedom of the macrospin state and the reconfigurable inter-element coupling we provide a toolset for: redirecting magnons around a 2D disk network whilst conserving power and phase, amplitude gating out-performing 1D gate designs\cite{haldar2016reconfigurable,gartside2020current,cramer2018magnon,wu2018magnon,cornelissen2018spin} and magnon phase-inversion. The fine control of spin-wave properties and tailoring of inter-element coupling provides a canvas for complex, reconfigurable, magnonic logic architectures shown \textit{via} a magnonic interferometer performing XNOR functionality. State-control is achieved through local writing \textit{via} a high-moment MFM tip\cite{gartside2018realization,gartside2016novel}. We demonstrate reconfigurable control of macrospin axis and vortex chirality and polarity, enabling full control of disk states across a range of dimensions. 



\section*{Results and Discussion}

\subsection*{Working principle of the state-control method}

\begin{figure}[h!]

\includegraphics[width=0.5\textwidth]{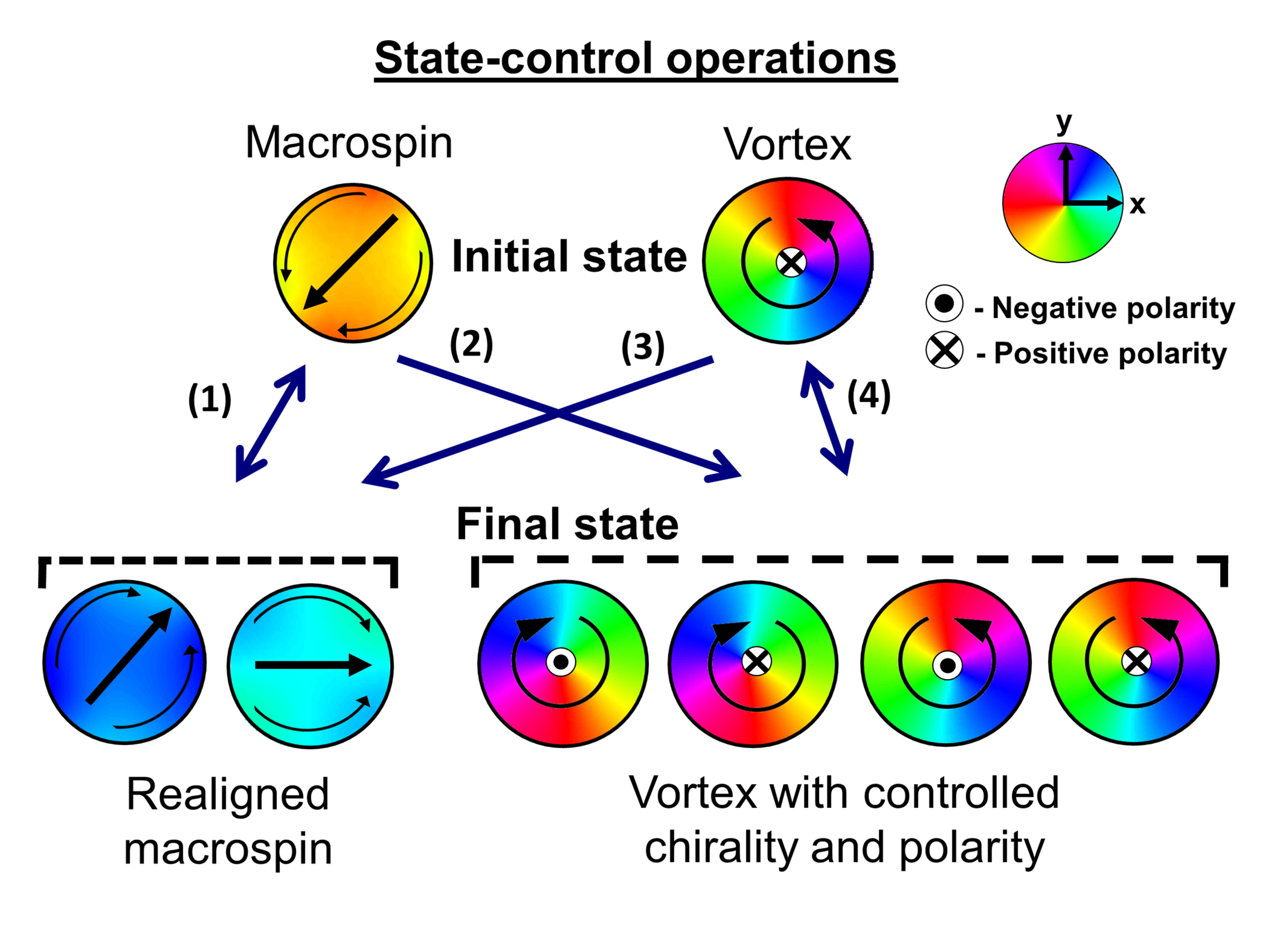}
\caption{Schematic of the four state-control operations. 1) Macrospin to realigned macrospin. 2) Macrospin to vortex with controlled polarity and chirality. 3) Vortex to realigned macrospin. 4) Vortex to another vortex with controlled chirality and polarity.}
\label{Fig0} \vspace{-1em}
\end{figure}

We present an RMC comprising ferromagnetic nanodisks with two metastable disk states: the macrospin state with large stray field from its uniform magnetisation and the vortex state with a closed circular magnetisation (clockwise (CW) or anticlockwise (ACW) chirality) with no stray field except at the vortex core where magnetisation points out-of-plane $\pm \hat{z}$ (sign determining vortex polarity). When controlling nanodisk states, we thus have four operations depending on the initial and final disk state as illustrated in Fig. \ref{Fig0}: 1) macrospin realignment, 2) macrospin to vortex with polarity and chirality control, 3) vortex to macrospin and 4) vortex to vortex with different polarity and/or chirality. The range of available control operations for a given nanodisk is dependent on the disk dimensions.

\begin{figure}[h!]
\centering

\includegraphics[width=\textwidth]{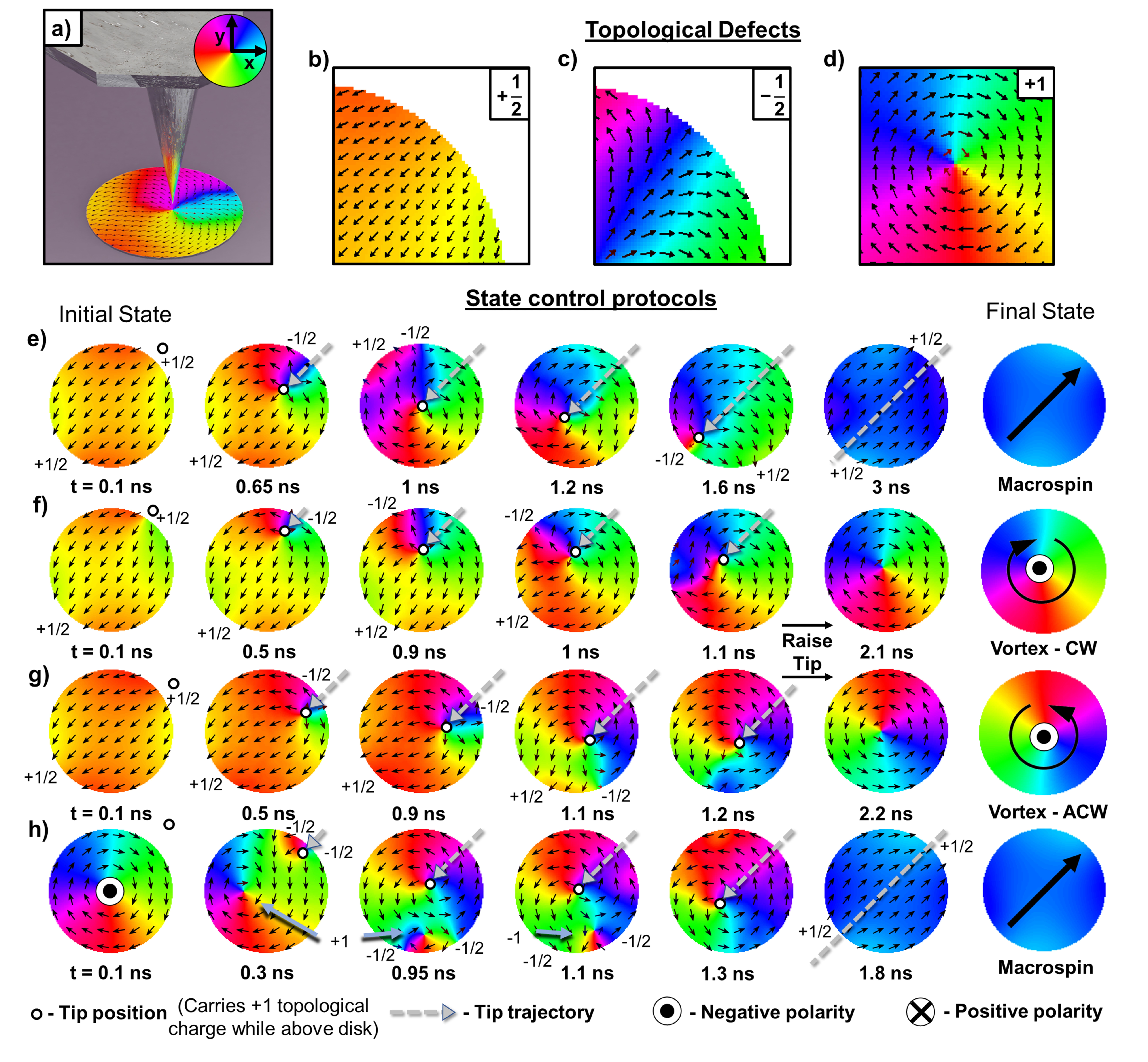}
\caption{a) Schematic of a magnetic force microscope (MFM) tip traversing above the surface of a magnetic nanodisk. The stray field from the tip causes the local distortion of the magnetisation texture. Colours represent the direction of magnetisation. b-d) Topological defect spin configurations along with their associated winding numbers describing the manner in which spins distort away from a collinear state in a thin-film ferromagnet. Time-evolution series showing four state-control scenarios; e) macrospin to realigned macrospin, f) macrospin to CW vortex, g) macrospin to ACW vortex, h) CW vortex to realigned macrospin.  }
\label{Fig1-Micromagnetics} \vspace{-1em}
\end{figure}

 Fig. \ref{Fig1-Micromagnetics} a) shows a schematic of the system comprising a high moment MFM tip traversing the surface of a ferromagnetic nanodisk (here permalloy (Py)). The stray field emanating from the tip (H\textsubscript{tip}) is modelled as monopolar\cite{magiera2012magnetic, gartside2016novel, gartside2018realization}, diverging away from the tip-apex. As an aid to understanding the writing process, spin configurations of topological defects \cite{tchernyshyov2005fractional,pushp2013domain} and their associated winding numbers are shown in Fig. \ref{Fig1-Micromagnetics} b-d). These are points of distortion within the magnetisation texture that cannot be smoothly unwound \cite{magiera2012magnetic}, injection and manipulation of which underpin the writing process. Time-evolution series of four different writing operations preparing macrospin and CW and ACW vortex states are presented in Fig. \ref{Fig1-Micromagnetics} e-h). Each series views a single 150 nm diameter $\times$ 10 nm thick Py nanodisk from above. Disk dimensions are selected for good vortex and macrospin metastability \cite{cowburn1999single,chung2010phase}. The tip starts at a distance of 400 nm from the disk in the \textit{xy} plane and traverses the disk at a height of 2.5 nm above the disk surface, simulating a contact AFM scan height\cite{gartside2016novel}. Videos of the time evolution are provided in the Supplementary Information.

\subsubsection*{Writing from a macrospin state}
As a tip carrying positive magnetic charge (\textit{i.e.} magnetised in the -$\hat{z}$ direction) approaches a macrospin nanodisk in the \textit{xy}-plane, H\textsubscript{tip} causes the entire macrospin state to rotate such the such that the magnetisation (M\textsubscript{disk}) orients away from the tip position (towards the tip for a tip carrying negative magnetic charge), minimising dipolar energy. Once the tip starts moving over the surface of the nanodisk [Fig. \ref{Fig1-Micromagnetics} e), t = 0.65 ns], the local magnetisation distorts away from the macrospin configuration to align with the radial H\textsubscript{tip} profile. This results in the injection of a vortex, which remains under the tip connected to the disk-edge by a region of reversed magnetisation (analogous to a domain wall) bound by the exchange-energy penalty required to realign the magnetisation. As the tip continues to move, the region of reversed magnetisation grows resulting in an increase in system energy. To reduce this energy, the reversed magnetisation region then rotates to minimise the exchange energy [Fig. \ref{Fig1-Micromagnetics} e), t = 1 ns], in the process bringing the disk to a vortex state [Fig. \ref{Fig1-Micromagnetics} e), t = 1.2 ns].

From here, there are two options; continue the tip movement across the surface of the disk, leaving a macrospin state [Fig. \ref{Fig1-Micromagnetics}. e)] or stop and raise the tip above the surface of the nanodisk, leaving a vortex state [Fig. \ref{Fig1-Micromagnetics} f,g)]. In the macrospin case, the macrospin orientation is decided by M\textsubscript{disk} aligning with the radial $x,y$ component of H\textsubscript{tip}. Fig. \ref{Fig1-Micromagnetics} has H\textsubscript{tip} extending radially outward so M\textsubscript{disk} aligns away from the final tip position, this is reversed by inverting the tip polarity. If instead the tip is stopped and raised above the centre of the disk, the magnetisation relaxes into a vortex state. Fig. \ref{Fig1-Micromagnetics} f,g) show that offsetting the tip-trajectory relative to the central disk axis gives a shorter pathway around one disk edge for spin-chain movement. The chain then traverses this pathway until the vortex is formed, with vortex chirality determined by the direction of spin-chain movement [Fig. \ref{Fig1-Micromagnetics} f) t = 1.1 ns, g) t = 1.2 ns].  The polarity of the vortex core is determined by the $\pm$\textit{z} magnetisation of the tip, allowing preparation of all five metastable disk states.

\begin{figure}
\includegraphics[width=0.5\textwidth]{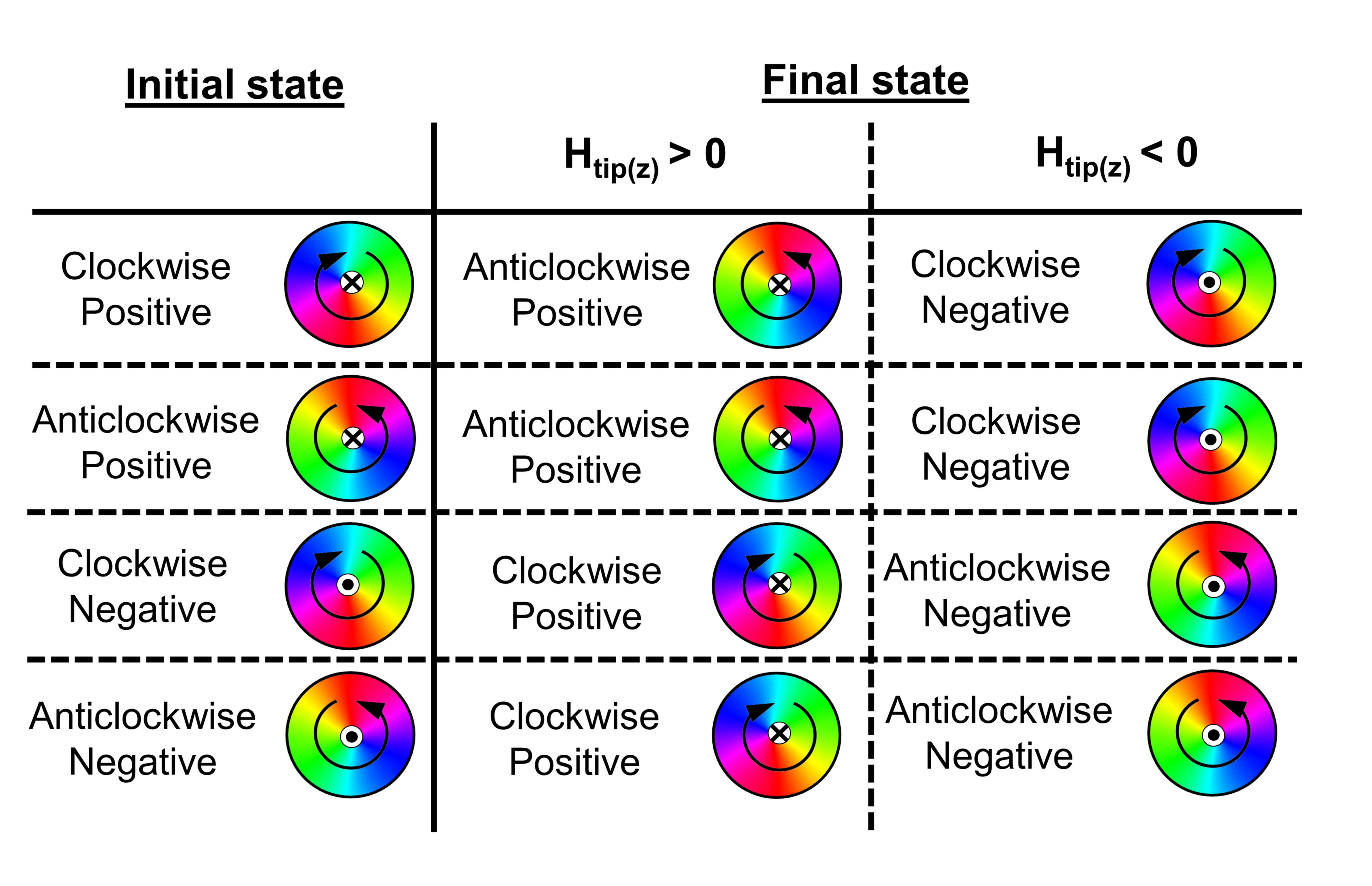}
\caption{Resulting microstate for each combination of vortex chirality, polarity and tip polarity when scanning, stopping and raising the tip above the nanodisk surface. If the initial vortex polarity aligns with (against) H\textsubscript{tip(z)}, the resulting chirality is anticlockwise (clockwise). The vortex polarity always aligns with the tip polarity.}
\label{Fig-vortex} \vspace{-1em}
\end{figure}

\subsubsection*{Writing from a vortex state}
Starting from a vortex state [Fig. \ref{Fig1-Micromagnetics}. h), Fig. \ref{Fig-vortex}], there are four possible initial vortex configurations and two tip polarities to consider. In all cases, scanning the tip across the entire surface of the nanodisk results in a macrospin state [Fig. \ref{Fig1-Micromagnetics} h)]. If we stop the scan while above the disk and raise the tip to produce a vortex, the resulting chirality and polarity depends on the polarity of both the initial vortex core and the tip. If the initial vortex polarity aligns with (against) H\textsubscript{tip(z)}, the resulting chirality is anticlockwise (clockwise). The resulting vortex-core polarity always aligns with tip polarity (Fig. \ref{Fig-vortex}). This provides a mechanism for full control over polarity and chirality of vortex states with a maximum of two operations (\textit{e.g.} going from clockwise positive to anticlockwise negative requires an intermediate clockwise positive state). In each case, the resulting dynamics depend on both the initial disk chirality and tip polarity. The dynamics of each vortex starting state are shown in the Supplementary Information along with a more technical description of the state-control methods in terms of topological defect control \cite{gartside2016novel,gartside2018realization,gartside2020current}.

\subsection*{Regimes of Control}

\begin{figure}[h!]
\centering

\includegraphics[width=\textwidth]{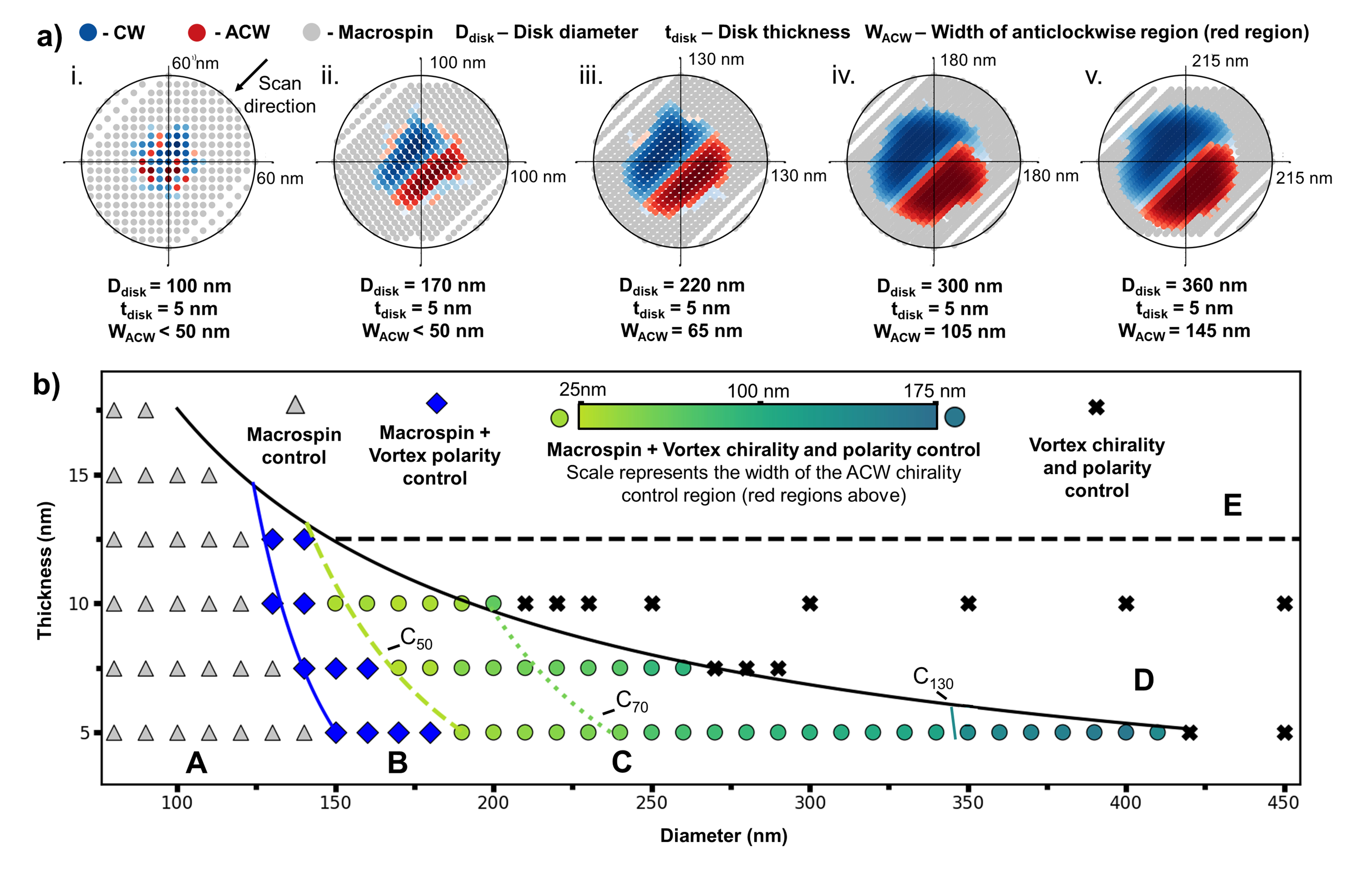}
\caption{a) Microstate-selection plots showing the resulting microstate after stopping and raising the tip above a range of nanodisk diameters (D\textsubscript{disk}). The tip travels in the -$\hat{x}$ -$\hat{y}$ direction. Each point in the diagram represents the point at which the tip stops before a being raised. In each case, the width of the anticlockwise region (W\textsubscript{ACW}) is labelled.
b) Microstate control phase diagram showing five distinct regimes: a macrospin control region (\textbf{A}, grey triangles) capable of defining macrospin direction, a macrospin and vortex polarity control region (\textbf{B}, blue squares), a complete control region capable of reconfiguring macrospin direction as well as vortex chirality and polarity (\textbf{C}, green circles) where the colour scale represents the width of the anticlockwise retraction-point region (\textit{i.e.} red region in a)), a vortex chirality and polarity control region without access to macrospin states (\textbf{D}, black crosses) which extends to disk diameters of at least 450 nm. Finally, a vortex polarity control region (\textbf{E}). The upper limit of this regime is greater than a disk thickness of 40 nm for all disk diameters presented. Also shown are three green, dashed lines representing an ACW region width of 50 nm (C\textsubscript{50}), 70 nm (C\textsubscript{70}) and 130 nm (C\textsubscript{130}).  \\
}
\label{Fig2-phase-diagram} \vspace{-1em}
\end{figure}

To study distinct state control regimes, the above protocols were performed on a range of disk dimensions. For each dimension, the tip is scanned across the disk surface, stopped above a certain point and raised for a sufficient time so that the magnetisation relaxes in the absence of H\textsubscript{tip}. All simulations are performed with H\textsubscript{tip} diverging away from the tip (H\textsubscript{tip(z)} $<$ 0).
Fig. \ref{Fig2-phase-diagram} a) i.-v. shows microstate-selection plots for a range of nanodisk diameters. Each circular point represents the location where the tip is halted and raised. The nanodisk ground state magnetisation depends on the competition between the demagnetisation energy favouring a divergence-free state and the exchange energy favouring parallel alignment of spins. The boundary between vortex and macrospin states has previously been calculated analytically\cite{jubert2004analytical} and approximately follows a t\textsubscript{disk} $\propto$ $1/{D\textsubscript{disk}}$ dependence where t\textsubscript{disk} and D\textsubscript{disk} are the disk thickness and diameter respectively\cite{cowburn1999single,chung2010phase}. Furthermore, local minima in the energy-landscape allows for metastable macrospin and vortex states across a range of disk dimensions \cite{metlov2008map,ostman2014hysteresis}.

Fig. \ref{Fig2-phase-diagram} b) shows a microstate-control phase diagram comprising five distinct regimes (\textbf{A}-\textbf{E}) which can be used to select the desired functionality of the writing protocol. At small disk dimensions, the macrospin state is energetically favoured [Fig. \ref{Fig2-phase-diagram} a.i), Fig. \ref{Fig2-phase-diagram} b) regime \textbf{A}]. In this regime it is possible to control the direction of M\textsubscript{disk} but whilst metastable vortex states are accessible in simulation, the required precision to realise a vortex exceeds what is experimentally achievable hence vortex preparation is unfeasible. The boundary of this region shows a similar dimension dependence to the theoretical lower limit of vortex states\cite{cowburn1999single}. On the other hand, for larger disk diameters and thicknesses [Fig. \ref{Fig2-phase-diagram} b) regime \textbf{D}], macrospins are no longer stable. In this regime, if t\textsubscript{disk} $\leq$ 10 nm, control of vortex polarity and chirality is still possible [Fig. \ref{Fig-vortex}]. The lower bound of this regime closely follows a $1/{D\textsubscript{disk}}$ dependence. For greater thicknesses, vortex polarity may still be controlled [Fig. \ref{Fig2-phase-diagram} b) regime \textbf{E}]. In this regime, H\textsubscript{tip} no longer induces a vortex. Instead, the stray field of the tip injects a singularity (bloch point) which propagates through the thickness of the disk resulting in vortex core reversal \cite{thiaville2003micromagnetic}. The upper limit of this regime is t\textsubscript{disk} > 40 nm for all diameters presented. Here we use a conservative tip strength value chosen to correspond to commercially available high-moment tips. If a higher tip strength is chosen the upper limit of regimes \textbf{D} and \textbf{E} would increase, allowing further control of vortex states.

Fig. \ref{Fig2-phase-diagram} b) regimes \textbf{B} and \textbf{C} represent the nanodisk dimensions where both macrospin and vortex states are metastable.  In regime \textbf{B}, the retraction-point area for vortex states exceeds 50 nm allowing preparation of macrospin and vortex states with controlled macrospin axis and vortex polarity. In regime \textbf{C}, clear, experimentally achievable retraction-point regions to prepare clockwise and anticlockwise vortex states emerge [Fig. \ref{Fig2-phase-diagram} a) iii-iv.] allowing control over all five metastable states.
This complete-control region spans a range of disk dimensions where the retraction-point regions for each vortex chirality (red and blue regions in Fig. \ref{Fig2-phase-diagram} a)) increase with increasing disk diameter. If instead H\textsubscript{tip} converges towards the tip point (H\textsubscript{tip(z)} $>$ 0), the microstate control phase diagram remains the same however the chirality is opposite to that presented in Fig. \ref{Fig2-phase-diagram} a). Here, the tip travels in the -$\hat{x}$ -$\hat{y}$ direction. If a different direction is chosen, the microstate-selection plots in Fig. \ref{Fig2-phase-diagram} a) rotate accordingly. The technique retains full functionality when the tip is lowered into dense nanodisk arrays (extended discussion of this case is provided in the supplementary information) allowing tailoring of inter-disk coupling \textit{via} active (macrospin) and inactive (vortex) state selection\cite{bhanja2016non} or vortex polarity control \cite{kumar2014magnetic,jung2011tunable}, tuning spin-wave emission in stacked vortices \textit{via} chirality control\cite{wintz2016magnetic} and injection of magnetic skyrmions\cite{ognev2020magnetic,zheng2017direct,zeissler2018discrete} without disturbing surrounding magnetic states.

\subsection*{Nanodisk-based reconfigurable magnonic crystal}

Utilising the state-control tools established above, we present an RMC offering diverse functionality with no reliance on global magnetic field. The design comprises a 2D array of ferromagnetic nanodisks (here Py) with a diameter of 150 nm, height of 5 nm and inter-disk spacing of 25 nm, within control regime \textbf{B} affording macrospin and vortex state control [Fig. \ref{Fig2-phase-diagram} b)]. Nanodisk dimensions were chosen to both optimise RMC performance and allow preparation of macrospin and vortex states. 

\subsubsection*{Magnon waveguiding and gating}

\begin{figure}[htbp!]
\centering

\includegraphics[width=\textwidth]{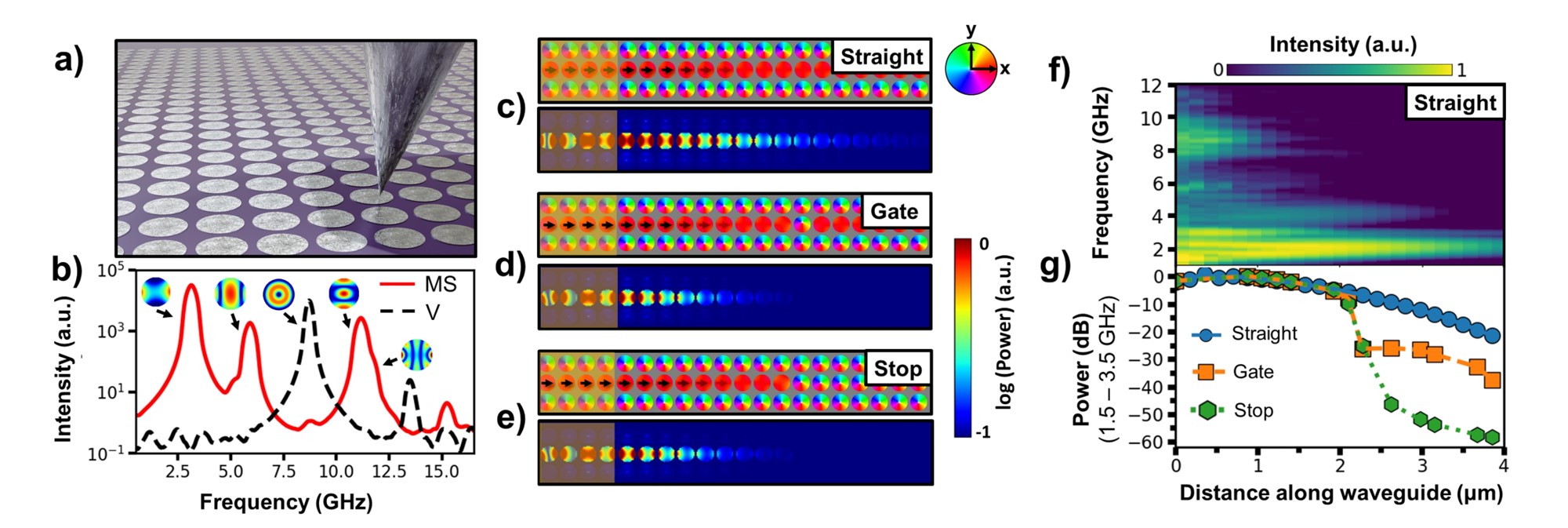}
\caption{a) Schematic of the 2D RMC comprising densely packed nanodisks. b) Spin-wave spectra of a 1D nanodisk array with a disk width of 150 nm, height of 5 nm and inter-disk separation of 25 nm in macrospin (MS) and vortex (V) states excited by a out-of-plane sinc pulse exciting modes up to 25 GHz. The macrospin spectra shows three main resonance modes corresponding to the macrospin disk edge-mode (2.5 - 4.5 GHz), bulk-mode (5 - 7 GHz) and two high frequency modes (10 - 12.5 GHz). The vortex spectra contains one distinct radial mode (7.5 - 10 GHz). c-e) Three microstates and power spectra of the lowest frequency macrospin edge-mode showing spin-wave transmission along a macrospin pathway 'Straight' and blocking of transmission when introducing vortex states ('Gate', 'Stop'). f) Spin-wave spectra as a function of distance for a straight macrospin pathway. g) Power of the lowest frequency macrospin edge-mode (1.5 - 3.5 GHz) along the length of the waveguide for the 'Straight', 'Gate' and 'Stop' microstates shown in c-e). In each case the macrospin pathway acts as a magnonic waveguide for the macrospin edge modes.  In each simulation the first 4 columns of disks (shaded gold) are excited with an out-of-plane sinc field exciting frequencies up to 25 GHz to model excitation \textit{via} a co-planar waveguide.}
\label{Fig3} \vspace{-1em}
\end{figure}

Fig. \ref{Fig3} b) shows the spin-wave spectra and spatial mode power profiles of a 1D array of macrospin (MS) and vortex (V) states excited with an out-of-plane H\textsubscript{ext} sinc-pulse capable of exciting modes up to 25 GHz, modelling excitation \textit{via} a coplanar waveguide. Vortex disks exhibit one significant mode corresponding to the n = 1 radial standing mode. The macropsin state exhibits a range of modes corresponding to the edge-mode (2.5 - 4.5 GHz), bulk-mode (5 - 7.5 GHz) and two high frequency modes (10 - 12.5 GHz)\cite{guo2013spectroscopy,PhysRevLett.107.127204}. It is important to note that the curved nature of the magnetisation and the breakdown of the macrospin approximation leads to the presence of distinct modes. Fig. \ref{Fig3} c-e) show magnetisation configurations and corresponding spatial power profiles for three RMC microstates, `Straight' [Fig. 
\ref{Fig3} c)], `Gate' [Fig. \ref{Fig3} d)] and `Stop' [Fig. \ref{Fig3} e)]. 

Fig. \ref{Fig3} f) shows the straight spin-wave spectra (averaged over each disk) along the waveguide length. In all cases, the central macrospin line is magnetised in the \textit{M} = +\^{x} direction and the first four columns of disks are excited (highlighted gold). The macrospin disks act as a waveguide, enabling magnon power transfer \textit{via} strong inter-macrospin disk dipolar coupling and a relative lack of dipolar coupling and shift in mode frequency to vortex disks. Fig. \ref{Fig3} c) shows the spin-wave spectra along the length of the waveguide showing magnon propagation occurs chiefly in the 1.5 - 3.5 GHz range, with the edge-mode power transmission for the three RMC microstates presented in Fig. \ref{Fig3} g).  Introducing a single vortex disk along the macrospin pathway ['Gate', Fig. \ref{Fig3} d)] or ending the pathway ['Stop', Fig. \ref{Fig3} e)] blocks power transmission with "on/off" ratios  of $\sim$ 30 and $\sim$ 10\textsuperscript{4} respectively. Therefore, amplitude gating is achieved by selecting a particular nanodisk to be a macrospin (gate "off"), allowing power transmission, or a vortex (gate "on"), blocking transmission. "On/off" ratios are calculated from the ratio of integrated intensity between the gated and ungated case at the 19\textsuperscript{th} disk from left RMC edge. These ratios match or out-perform previous gate 1D gate designs \cite{haldar2016reconfigurable,gartside2020current,cramer2018magnon,wu2018magnon,cornelissen2018spin} due to the shifted mode frequency and reduced dipolar coupling between macrospin and vortex states.

\begin{figure}[h!]
\centering

\includegraphics[width=\textwidth]{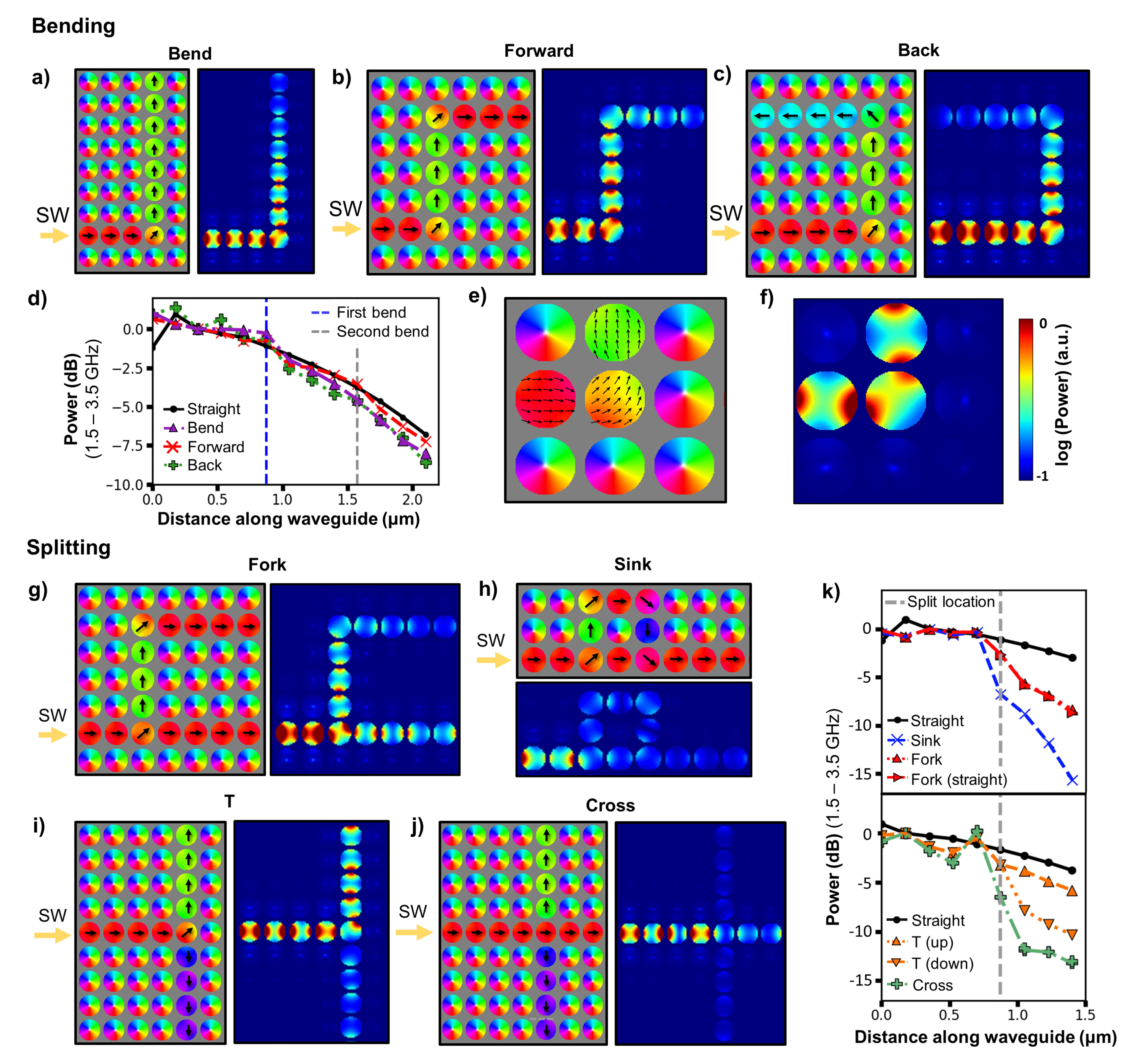}
\caption{Microstates and power spectra demonstrating bending of magnons across a 2D array for a a) single bend b) forward bend and c) backward bend. d) Power profiles for the various bend configurations compared to a straight macrospin waveguide that ran for an equivalent time (8 ns). All power profiles follow a similar trend to the straight case with small raises in power at each bend. e) Magnetisation profile and f) power spectra of a bend. Curling of the magnetisation at bend locations results in continued power distribution along the waveguide. g-j) Microstates and power spectra and demonstrating distribution of magnons across a 90$^{\circ}$ split (`Fork'), splitting and rejoining (`Sink') and two 180$^{\circ}$ split configurations (`T', `Cross'). k) Each state is compared to a straight macrospin waveguide that ran an equivalent time (8 ns).  For a 90\textsuperscript{$\circ$} split (`Fork') the power is distributed equally along both pathways with a 5 dB drop along each pathway. If the power is split and rejoined (`Sink'), destructive interference results in a 12 dB power drop and thus acts as a power sink. For a 180\textsuperscript{$\circ$} split (`T') the power splits unevenly with a preferential direction depending on the direction of magnetisation of the central disk. The addition of two or more macrospin disks placed along the straight pathway (`Cross') causes the central disk to relax along the x-direction, removing the preferential splitting. In each case, two macrospin disks further up the waveguide are excited with a sinusoidal field with \textit{f} = 2.3 GHz to mimic incoming spin-waves.}
\label{Fig4} \vspace{-1em}
\end{figure}



\subsubsection*{Shaping magnon pathways}

Nanodisk magnetisation is free to rotate in the XY-plane, well-suited to redirecting information across a 2D network. Macrospin pathways incorporating single- and double-bends were investigated, shown in Fig. \ref{Fig4} a-c). Two macrospin nanodisks at the left RMC edge are excited with a \textit{f} = 2.3 GHz sinusoidal field. Fig. \ref{Fig4} d) shows introducing bends to the macrospin pathway gives only a small magnon power loss, with a slight power increase at bend locations due to constructive interference between forward-propagating and backscattered spin-waves. The power distribution contrasts previous results where substantial power losses are incurred while bending\cite{garcia2015narrow,haldar2016reconfigurable} and is due to curling of the corner-disk magnetisation at bend locations [Fig. \ref{Fig4} e,f)] allowing excited edge-modes to remain close, providing a powerful and efficient method of redirecting magnons across a 2D array.

Another advantage of the freely-rotating nanodisk magnetisation is the ability to split magnon power across multiple pathways. Fig. \ref{Fig4} g-j) show several microstates performing magnon power distribution. If the macrospin pathway forms a 90\textsuperscript{$\circ$} split [Fig. \ref{Fig4} g), `Fork'], power is transmitted equally along both pathways due to the corner disk magnetisation relaxing at 45$^{\circ}$ to the original path, symmetrical to both forward branches with a relative 5 dB drop in each branch. If this split is then redirected back onto the original path \textit{via} a second bend [Fig. \ref{Fig4} h), `Sink'], a -12 dB power sink is achieved due to the destructive interference. If a 180\textsuperscript{$\circ$} split [Fig. \ref{Fig4} i), `T'] is introduced, the junction disk relaxes at 45$^{\circ}$ to either the `up' or `down' branch, breaking symmetry and resulting in preferential transmission along that branch ($\Delta$P between the two branches = 4.5 dB). By extending the macrospin pathway along the central axis by two or more macrospin disks [Fig. \ref{Fig4} j), `Cross'], the junction disk now remains aligned to the original pathway (here +\^{x} direction), restoring symmetry and allowing equal power distribution along `up' and `down' branches with a $\sim$ 9 dB reduction in each arm relative to the straight case. If the macrospin pathway is only extended by one macrospin disk, the junction disk remains asymmetric resulting in preferential power transmission. In all cases of bending and splitting, both spin-wave frequency and phase are conserved in all pathways. Combining these designs with gates allows for freedom in computational architecture enabling parallel processing of magnon information and the realisation of magnonic multiplexers capable of selecting information from multiple input lines to one output line.

\begin{figure}[h!]
\centering

\includegraphics[width=\textwidth]{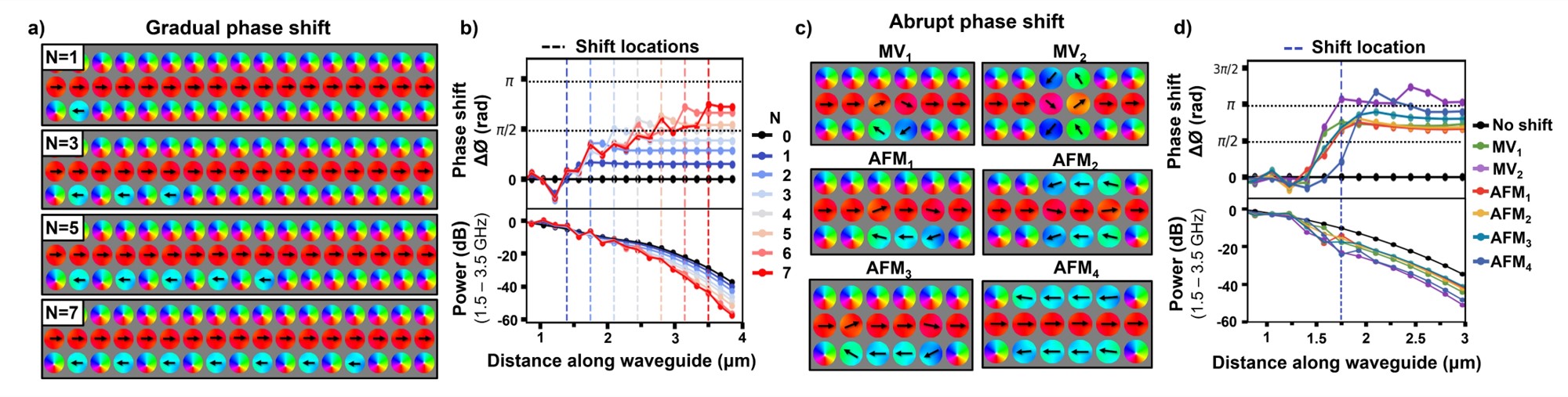}
\caption{a) Microstates of four states with increasing number of anti-parallel macrospin disks interspersed by vortex disks achieving incremental shifting of spin-wave phase.  b) Spin-wave phase shift and power distribution along the waveguide for the states in a) ranging from macrospin disk number N = 0 - 7. c) Multiple microstates (MV\textsubscript{1,2} and AFM\textsubscript{1,2,3,4}) used to achieve an abrupt phase shift.  d) Corresponding phase shift and power profiles as a function of distance along the waveguide. In b) and d) the dashed line represents the position of the phase shifting disks. In all cases, magnon phase is calculated by comparing \textit{m\textsubscript{z} \textit{vs} t} to the `no shift' case.}
\label{Fig5} \vspace{-1em}
\end{figure}

\subsubsection*{Phase control}

Fig. \ref{Fig5} a,b) show that by introducing macrospin disks adjacent to the waveguide, a gradual phase shift is induced in the travelling spin-wave. The magnitude of the phase shift depends on the number of adjacent disks increasing to 0.7 $\pi$ rad for seven nanodisks. Fig. \ref{Fig5} c,d) show that a more abrupt phase inversion is achieved by adding clusters of macrospin disks adjacent to waveguide. In each case, the adjacent macrospin disks form either  `macrovortex' (MV\textsubscript{1,2}) or  `antiferromagnetic' (AFM\textsubscript{1,2,3,4}) magnetisation profiles. Here, phase shifts of up to $\pi$ rad are achieved. In all cases, magnon power is transferred to the adjacent macrospin disks causing a relative power loss. This range of microstates is non-exhaustive and demonstrates how the state-control method and RMC design allow for fine, reconfigurable magnonic phase control.

\subsubsection*{All-magnonic interferometry}

\begin{figure}[h!]
\centering

\includegraphics[width=\textwidth]{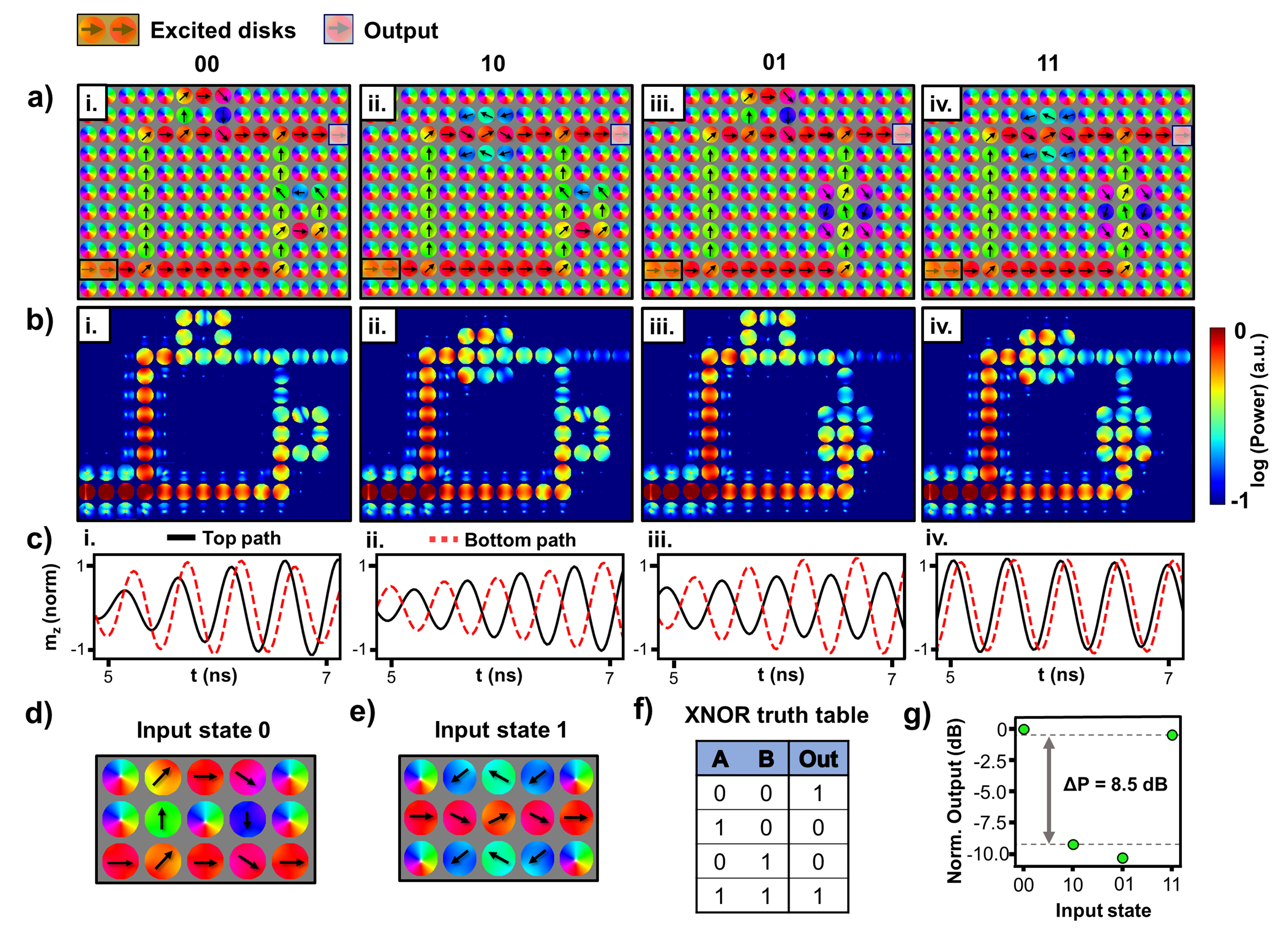}
\caption{a) i.-iv. Microstates and b) i.-iv. power spectra of four interferometer inputs. The first two disks on the macrospin pathway are excited with an out-of-plane sinusoidal field with \textit{f} = 2.3 GHz. The output is taken as the power on the final disk along the macrospin pathway. c) i.-iv. Shows m\textsubscript{z} \textit{vs} t of the top (solid black line) and bottom (red dashed line) pathways at the disks just before recombination.  Microstates of d) input state 0 (power sink) and e) logic state 1 (phase shift).  f) XNOR logic truth table showing input and output values. g)  Output power of the final macrospin disk between 1.5 - 3.5 GHz at the last 2 ns of the simulation to allow sufficient time for spin-waves to propagate through both pathways (normalised to the 00 case). }
\label{Inter} \vspace{-1em}
\end{figure}

The methods described above provide a powerful toolset for manipulation of magnon amplitude and phase, permitting more complex architectures for wave-based computation. We demonstrate this here \textit{via} an all-magnonic interferometer, integrating the designs described above to achieve XNOR logic functionality. 
Depicted in Fig. \ref{Inter}, each logic state contains a combination of power sink branches representing an input of 0 [Fig. \ref{Inter} c)] and phase shift branches representing an input of 1 [Fig. \ref{Inter} d)]. The inclusion of the power sink is necessary to compensate for the power loss induced by the phase shift. When both paths contain a power sink (00) or a phase shift (11) the two pathways meet with matching phase and constructively interfere resulting in a high output power. If the inputs of the two arms differ, the spin-waves meet with a $\sim$ $\pi$ phase shift, resulting in destructive interference and low power output [Fig. \ref{Inter} f)]. The 1 and 0 output states differ by a power of $\sim$8.5 dB [Fig. \ref{Inter} f)], allowing clear differentiation. More detail into the design considerations for complex computational microstates can be found in the Supplementary Information.


\section*{Conclusion}
In this work, we propose a nanodisk-based reconfigurable magnonic crystal where tailoring the coupling between neighbouring elements is employed to achieve gating, steering and phase shifting of spin-waves across a 2D network. Through microstate engineering, we demonstrate a powerful, flexible platform for hosting next-generation magnonic technologies with substantial attractive benefits over existing nanowire-based RMC designs. The interferometry demonstrated here represents an initial proof of concept of the efficacy and value of disk-based RMCs, with promise of deeper and more complex functionality for wave and neuromorphic computation as disk-based systems are further explored. This is made possible by the reconfigurable state-control \textit{via} high-moment MFM tip, highlighting the continued utility of local microstate control techniques \cite{gartside2016novel,gartside2018realization}. Through use of self-biasing elements, the design requires no additional energy once the states are initialised promising low power, low heat computation.

Whilst the design cannot currently compete with mature CMOS technologies, the proof-of-concept properties show promise for parallel processing and interference-based computation which are not well catered for by CMOS. As methods to translate control from a scanning tip to on-chip circuitry evolve \cite{gartside2020current,nance2020spin}, the results presented here may offer opportunities to evolve scalable solid-state technology capable of combined memory and computational functionality, overcoming another key bottleneck of current computational architecture.

\subsection*{Methods}
\subsubsection*{State-control method simulations}
All simulations for the state-control method were performed using  the object-oriented micromagnetic framework (OOMMF)\cite{donahue1999oommf}. Magnetic parameters for Permalloy (Ni\textsubscript{80}Fe\textsubscript{20}) of  M\textsubscript{sat}  = 860 $\times$ 10\textsuperscript{3}   kA/m, A = 13$\times$10\textsuperscript{-12} Jm\textsuperscript{-1}, zero magnetocrystalline anisotropy and a Gilbert damping parameter, $\alpha$ = 0.01 are used. The point probe approximation stating that an MFM tip may be described by a point monopole moment at small tip–sample separations\cite{hartmann1989point,lohau1999quantitative} was used. This approximation is widely used in MFM simulations, and previous work analysing systems with similar dynamics has shown similar results \cite{magiera2012magnetic,magiera2014magnetic,gartside2016novel,gartside2018realization}.

The simulated disks were discretized into 2.5 nm $\times$ 2.5 nm $\times$ 2.5 nm cells, smaller than the magnetostatic exchange length, l\textsubscript{s} $= \sqrt{2A_{ex}/\mu_{0}M_{sat}^2}$ = 5.29 nm. A preliminary investigation using 1 nm $\times$ 1 nm $\times$ 1 nm cell sizes returned equivalent results.

The field from the MFM tip was modelled as monopolar, producing a radial field $H = \frac{\mu_0}{4\pi}\frac{q_{tip}}{r^2}$ at a distance r from the charge. During the simulation the magnetic charge moved 1 nm every 20 ps giving a velocity of 50 m/s. This is faster than experimental speeds of 10\textsuperscript{-4} m/s but slow enough to avoid any dynamic excitation. In each case the disk is first magnetised with M\textsubscript{disk} pointing away from (towards) the positively (negatively) charged tip. This is done to allow complete M\textsubscript{disk} relaxation before H\textsubscript{tip} starts to influence the magnetisation.  In each simulation, the tip starts at a distance of 400 nm from the disk centre (in the plane of the disk) to allow M\textsubscript{disk} to relax before the sample-tip interaction takes place. The charge travels in the -$\hat{x}$ -$\hat{y}$ direction. This was chosen arbitrarily and the effect of scan direction does not effect the final results. 

For the microstate control phase diagram, disks with dimensions up to 500 nm in width and 40 nm in height were simulated. Larger dimensions were not possible due to the large computation time needed. As such, the upper bound of the polarity control region [Fig. \ref{Fig2-phase-diagram} b), region \textbf{E}] is unknown.

\subsubsection*{2D RMC simulations}
Simulations were performed using MuMax\cite{vansteenkiste2014design}. The same magnetic parameters were used as in the state-control simulations with the exception of the Gilbert damping parameter which is reduced to $\alpha$ = 0.006 for closer correspondence with previous experimental studies. The simulated disks were discretized into 2.5 nm $\times$ 2.5 nm $\times$ 2.5 nm cells. In all simulations, the microstate was relaxed before excitation. Nanodisk dimensions were chosen to both optimise RMC performance as well as be in a regime capable of supporting uniform and macrospin states. In order to prevent backscattering at the end of the arrays, a spin-wave absorber with $\alpha$ following a parabolic profile was introduced at the end of the macrospin pathway\cite{venkat2018absorbing}.

The macrospin and vortex spectra obtained in Fig. \ref{Fig3} b) ran for a total of 20 ns and the magnetisation is saved every 5 ps. Periodic boundary conditions extending by 5 unit cells along the x-axis were used to mimic a 1D array. The disk is then excited with an out-of-plane sinc-pulse with a cutoff frequency of 25 GHz. The spectra and power-profiles are them obtained by performing an FFT on each cell and summing the vector magnitudes of the fourier components. The remainder of the simulations saved the magnetisation in 20 ps time-steps. For the Straight, Gate and Stop cases in Fig. \ref{Fig3}, an out-of-plane sinc pulse H\textsubscript{ext} exciting modes up to 25 GHz was applied to the first four columns of nanodisks to replicate excitation \textit{via} a co-planar waveguide. These simulations ran for a total of 12 ns. For the remaining simulations, the first two macrospin disks were excited with an out-of-plane sinusoidal field with a frequency \textit{f} = 2.3 GHz to replicate an incoming travelling magnon. The bending, splitting and phase shift simulations in Fig. \ref{Fig4}, \ref{Fig5}, \ref{Inter} each ran for 8 ns to provide comparable results. The all-magnonic interferometery simulations ran for a total of 12 ns. The power was taken from the integrated intensity of the final macrospin disk over the last 2 ns of the simulation to allowing sufficient time for spin-waves to travel both pathways and interfere and to avoid any initial transients in the oscillation amplitude. The combination of power sink and phase shift cluster was chosen to optimise the interferometer performance. The length of the simulation times has been verified in the Supplementary Information.

In all cases, the vortex chirality and polarity were set as clockwise and positive. An investigation into varying chirality and polarity revealed no change in magnon properties as expected due to the weak dipolar coupling of vortex states.

\subsection*{Author Contributions}
KDS, JCG, AV and WRB conceived the work.
\\
JCG wrote the initial code for the simulation of the state-control protocols. 
\\
TD wrote the initial code the simulations of the 2D RMC.
\\
DMA wrote the initial code for generation of spin-wave spectra.
\\
AV assisted with the design of microstates for magnon splitting
\\
KDS further developed and expanded the simulation codes, performed all of the simulations and analysis, drafted the manuscript and designed and rendered CGI visuals.
\\
All authors contributed to discussion and refinement of the manuscript.

\subsection*{Acknowledgements}
This work was supported by the Leverhulme Trust (RPG-2017-257) to WRB. 
TD and AV were supported by the EPSRC Centre for Doctoral Training in Advanced Characterisation of Materials (Grant No. EP/L015277/1). Simulations were performed on the Imperial College London Research Computing Service\cite{hpc}. The authors would like to thank L. F. Cohen of Imperial College London for enlightening discussion and comments.\\

\noindent
\textbf{Supporting Information Available:} Animations of state-control protocols. A description of state-control protocols in terms of topological defects. State-control in non-ideal disks. State-control in nanodisk arrays. A comparison of excitation methods used in the RMC simulations. Further design considerations when splitting magnons across 180$^{\circ}$. Design considerations for logic functionality. Verification of RMC simulation times.  This material is available free of
charge \textit{via} the Internet at http://pubs.acs.org

\label{Bibliography}
\bibliographystyle{apsrev4-1}
\bibliography{sample4.bib}

\providecommand{\latin}[1]{#1}
\makeatletter
\providecommand{\doi}
  {\begingroup\let\do\@makeother\dospecials
  \catcode`\{=1 \catcode`\}=2 \doi@aux}
\providecommand{\doi@aux}[1]{\endgroup\texttt{#1}}
\makeatother
\providecommand*\mcitethebibliography{\thebibliography}
\csname @ifundefined\endcsname{endmcitethebibliography}
  {\let\endmcitethebibliography\endthebibliography}{}
\begin{mcitethebibliography}{99}
\providecommand*\natexlab[1]{#1}
\providecommand*\mciteSetBstSublistMode[1]{}
\providecommand*\mciteSetBstMaxWidthForm[2]{}
\providecommand*\mciteBstWouldAddEndPuncttrue
  {\def\EndOfBibitem{\unskip.}}
\providecommand*\mciteBstWouldAddEndPunctfalse
  {\let\EndOfBibitem\relax}
\providecommand*\mciteSetBstMidEndSepPunct[3]{}
\providecommand*\mciteSetBstSublistLabelBeginEnd[3]{}
\providecommand*\EndOfBibitem{}
\mciteSetBstSublistMode{f}
\mciteSetBstMaxWidthForm{subitem}{(\alph{mcitesubitemcount})}
\mciteSetBstSublistLabelBeginEnd
  {\mcitemaxwidthsubitemform\space}
  {\relax}
  {\relax}

\bibitem[Khitun \latin{et~al.}(2010)Khitun, Bao, and Wang]{khitun2010magnonic}
Khitun,~A.; Bao,~M.; Wang,~K.~L. Magnonic Logic Circuits. \emph{J. Phys. D:
  Appl. Phys.} \textbf{2010}, \emph{43}, 264005\relax
\mciteBstWouldAddEndPuncttrue
\mciteSetBstMidEndSepPunct{\mcitedefaultmidpunct}
{\mcitedefaultendpunct}{\mcitedefaultseppunct}\relax
\EndOfBibitem
\bibitem[Khitun and Wang(2011)Khitun, and Wang]{khitun2011non}
Khitun,~A.; Wang,~K.~L. Non-Volatile Magnonic Logic Circuits Engineering.
  \emph{J. Appl. Phys.} \textbf{2011}, \emph{110}, 034306\relax
\mciteBstWouldAddEndPuncttrue
\mciteSetBstMidEndSepPunct{\mcitedefaultmidpunct}
{\mcitedefaultendpunct}{\mcitedefaultseppunct}\relax
\EndOfBibitem
\bibitem[Chumak \latin{et~al.}(2015)Chumak, Vasyuchka, Serga, and
  Hillebrands]{chumak2015magnon}
Chumak,~A.~V.; Vasyuchka,~V.~I.; Serga,~A.~A.; Hillebrands,~B. Magnon
  Spintronics. \emph{Nat. Phys.} \textbf{2015}, \emph{11}, 453--461\relax
\mciteBstWouldAddEndPuncttrue
\mciteSetBstMidEndSepPunct{\mcitedefaultmidpunct}
{\mcitedefaultendpunct}{\mcitedefaultseppunct}\relax
\EndOfBibitem
\bibitem[Romera \latin{et~al.}(2018)Romera, Talatchian, Tsunegi, Araujo, Cros,
  Bortolotti, Trastoy, Yakushiji, Fukushima, Kubota, Yuasa, Ernoult,
  Vodenicarevic, Hirtzlin, Locatelli, Querlioz, and Grollier]{romera2018vowel}
Romera,~M.; Talatchian,~P.; Tsunegi,~S.; Araujo,~F.~A.; Cros,~V.;
  Bortolotti,~P.; Trastoy,~J.; Yakushiji,~K.; Fukushima,~A.; Kubota,~H.;
  Yuasa,~S.; Ernoult,~M.; Vodenicarevic,~D.; Hirtzlin,~T.; Locatelli,~N.;
  Querlioz,~D.; Grollier,~J. Vowel Recognition with Four Coupled Spin-Torque
  Nano-Oscillators. \emph{Nature} \textbf{2018}, \emph{563}, 230--234\relax
\mciteBstWouldAddEndPuncttrue
\mciteSetBstMidEndSepPunct{\mcitedefaultmidpunct}
{\mcitedefaultendpunct}{\mcitedefaultseppunct}\relax
\EndOfBibitem
\bibitem[Torrejon \latin{et~al.}(2017)Torrejon, Riou, Araujo, Tsunegi, Khalsa,
  Querlioz, Bortolotti, Cros, Yakushiji, Fukushima, Kubota, Yuasa, Stiles, and
  Grollier]{torrejon2017neuromorphic}
Torrejon,~J.; Riou,~M.; Araujo,~F.~A.; Tsunegi,~S.; Khalsa,~G.; Querlioz,~D.;
  Bortolotti,~P.; Cros,~V.; Yakushiji,~K.; Fukushima,~A.; Kubota,~H.;
  Yuasa,~S.; Stiles,~M.~D.; Grollier,~J. Neuromorphic Computing with Nanoscale
  Spintronic Oscillators. \emph{Nature} \textbf{2017}, \emph{547},
  428--431\relax
\mciteBstWouldAddEndPuncttrue
\mciteSetBstMidEndSepPunct{\mcitedefaultmidpunct}
{\mcitedefaultendpunct}{\mcitedefaultseppunct}\relax
\EndOfBibitem
\bibitem[Bhanja \latin{et~al.}(2016)Bhanja, Karunaratne, Panchumarthy, Rajaram,
  and Sarkar]{bhanja2016non}
Bhanja,~S.; Karunaratne,~D.~K.; Panchumarthy,~R.; Rajaram,~S.; Sarkar,~S.
  Non-Boolean Computing with Nanomagnets for Computer Vision Applications.
  \emph{Nat. Nanotechnol.} \textbf{2016}, \emph{11}, 177\relax
\mciteBstWouldAddEndPuncttrue
\mciteSetBstMidEndSepPunct{\mcitedefaultmidpunct}
{\mcitedefaultendpunct}{\mcitedefaultseppunct}\relax
\EndOfBibitem
\bibitem[Mizrahi \latin{et~al.}(2018)Mizrahi, Hirtzlin, Fukushima, Kubota,
  Yuasa, Grollier, and Querlioz]{mizrahi2018neural}
Mizrahi,~A.; Hirtzlin,~T.; Fukushima,~A.; Kubota,~H.; Yuasa,~S.; Grollier,~J.;
  Querlioz,~D. Neural-Like Computing with Populations of Superparamagnetic
  Basis Functions. \emph{Nat. Commun.} \textbf{2018}, \emph{9}, 1--11\relax
\mciteBstWouldAddEndPuncttrue
\mciteSetBstMidEndSepPunct{\mcitedefaultmidpunct}
{\mcitedefaultendpunct}{\mcitedefaultseppunct}\relax
\EndOfBibitem
\bibitem[Grollier \latin{et~al.}(2020)Grollier, Querlioz, Camsari,
  Everschor-Sitte, Fukami, and Stiles]{grollier2020neuromorphic}
Grollier,~J.; Querlioz,~D.; Camsari,~K.~Y.; Everschor-Sitte,~K.; Fukami,~S.;
  Stiles,~M.~D. Neuromorphic Spintronics. \emph{Nat. Electron.} \textbf{2020},
  1--11\relax
\mciteBstWouldAddEndPuncttrue
\mciteSetBstMidEndSepPunct{\mcitedefaultmidpunct}
{\mcitedefaultendpunct}{\mcitedefaultseppunct}\relax
\EndOfBibitem
\bibitem[Sangwan and Hersam(2020)Sangwan, and Hersam]{sangwan2020neuromorphic}
Sangwan,~V.~K.; Hersam,~M.~C. Neuromorphic Nanoelectronic Materials. \emph{Nat.
  Nanotechnol.} \textbf{2020}, 1--12\relax
\mciteBstWouldAddEndPuncttrue
\mciteSetBstMidEndSepPunct{\mcitedefaultmidpunct}
{\mcitedefaultendpunct}{\mcitedefaultseppunct}\relax
\EndOfBibitem
\bibitem[Grundler(2015)]{grundler2015reconfigurable}
Grundler,~D. Reconfigurable Magnonics Heats Up. \emph{Nat. Phys.}
  \textbf{2015}, \emph{11}, 438--441\relax
\mciteBstWouldAddEndPuncttrue
\mciteSetBstMidEndSepPunct{\mcitedefaultmidpunct}
{\mcitedefaultendpunct}{\mcitedefaultseppunct}\relax
\EndOfBibitem
\bibitem[Haldar \latin{et~al.}(2016)Haldar, Kumar, and
  Adeyeye]{haldar2016reconfigurable}
Haldar,~A.; Kumar,~D.; Adeyeye,~A.~O. A Reconfigurable Waveguide for
  Energy-Efficient Transmission and Local Manipulation of Information in a
  Nanomagnetic Device. \emph{Nat. Nanotechnol.} \textbf{2016}, \emph{11},
  437\relax
\mciteBstWouldAddEndPuncttrue
\mciteSetBstMidEndSepPunct{\mcitedefaultmidpunct}
{\mcitedefaultendpunct}{\mcitedefaultseppunct}\relax
\EndOfBibitem
\bibitem[Krawczyk and Grundler(2014)Krawczyk, and Grundler]{krawczyk2014review}
Krawczyk,~M.; Grundler,~D. Review and Prospects of Magnonic Crystals and
  Devices with Reprogrammable Band Structure. \emph{J. Phys.: Condens. Matter}
  \textbf{2014}, \emph{26}, 123202\relax
\mciteBstWouldAddEndPuncttrue
\mciteSetBstMidEndSepPunct{\mcitedefaultmidpunct}
{\mcitedefaultendpunct}{\mcitedefaultseppunct}\relax
\EndOfBibitem
\bibitem[Wang \latin{et~al.}(2017)Wang, Chumak, Jin, Zhang, Hillebrands, and
  Zhong]{wang2017voltage}
Wang,~Q.; Chumak,~A.~V.; Jin,~L.; Zhang,~H.; Hillebrands,~B.; Zhong,~Z.
  Voltage-Controlled Nanoscale Reconfigurable Magnonic Crystal. \emph{Phys.
  Rev. B} \textbf{2017}, \emph{95}, 134433\relax
\mciteBstWouldAddEndPuncttrue
\mciteSetBstMidEndSepPunct{\mcitedefaultmidpunct}
{\mcitedefaultendpunct}{\mcitedefaultseppunct}\relax
\EndOfBibitem
\bibitem[Topp \latin{et~al.}(2010)Topp, Heitmann, Kostylev, and
  Grundler]{topp2010making}
Topp,~J.; Heitmann,~D.; Kostylev,~M.~P.; Grundler,~D. Making a Reconfigurable
  Artificial Crystal by Ordering Bistable Magnetic Nanowires. \emph{Phys. Rev.
  Lett.} \textbf{2010}, \emph{104}, 207205\relax
\mciteBstWouldAddEndPuncttrue
\mciteSetBstMidEndSepPunct{\mcitedefaultmidpunct}
{\mcitedefaultendpunct}{\mcitedefaultseppunct}\relax
\EndOfBibitem
\bibitem[Neusser and Grundler(2009)Neusser, and Grundler]{neusser2009magnonics}
Neusser,~S.; Grundler,~D. Magnonics: Spin Waves on the Nanoscale. \emph{Adv.
  Mater.} \textbf{2009}, \emph{21}, 2927--2932\relax
\mciteBstWouldAddEndPuncttrue
\mciteSetBstMidEndSepPunct{\mcitedefaultmidpunct}
{\mcitedefaultendpunct}{\mcitedefaultseppunct}\relax
\EndOfBibitem
\bibitem[Kruglyak \latin{et~al.}(2010)Kruglyak, Demokritov, and
  Grundler]{kruglyak2010magnonics}
Kruglyak,~V.~V.; Demokritov,~S.~O.; Grundler,~D. Magnonics. \emph{J. Phys. D:
  Appl. Phys.} \textbf{2010}, \emph{43}, 264001\relax
\mciteBstWouldAddEndPuncttrue
\mciteSetBstMidEndSepPunct{\mcitedefaultmidpunct}
{\mcitedefaultendpunct}{\mcitedefaultseppunct}\relax
\EndOfBibitem
\bibitem[Chumak \latin{et~al.}(2017)Chumak, Serga, and
  Hillebrands]{chumak2017magnonic}
Chumak,~A.~V.; Serga,~A.~A.; Hillebrands,~B. Magnonic Crystals for Data
  Processing. \emph{J. Phys. D: Appl. Phys.} \textbf{2017}, \emph{50},
  244001\relax
\mciteBstWouldAddEndPuncttrue
\mciteSetBstMidEndSepPunct{\mcitedefaultmidpunct}
{\mcitedefaultendpunct}{\mcitedefaultseppunct}\relax
\EndOfBibitem
\bibitem[Vedmedenko \latin{et~al.}(2020)Vedmedenko, Kawakami, Sheka,
  Gambardella, Kirilyuk, Hirohata, Binek, Chubykalo-Fesenko, Sanvito, Kirby,
  Grollier, Everschor-Sitte, Kampfrath, You, and Berger]{vedmedenko20202020}
Vedmedenko,~E.~Y.; Kawakami,~R.~K.; Sheka,~D.; Gambardella,~P.; Kirilyuk,~A.;
  Hirohata,~A.; Binek,~C.; Chubykalo-Fesenko,~O.~A.; Sanvito,~S.; Kirby,~B.;
  Grollier,~J.; Everschor-Sitte,~K.; Kampfrath,~T.; You,~C.-Y.; Berger,~A. The
  2020 Magnetism Roadmap. \emph{J. Phys. D: Appl. Phys.} \textbf{2020}, \relax
\mciteBstWouldAddEndPunctfalse
\mciteSetBstMidEndSepPunct{\mcitedefaultmidpunct}
{}{\mcitedefaultseppunct}\relax
\EndOfBibitem
\bibitem[Gliga \latin{et~al.}(2020)Gliga, Iacocca, and
  Heinonen]{gliga2020dynamics}
Gliga,~S.; Iacocca,~E.; Heinonen,~O.~G. Dynamics of Reconfigurable Artificial
  Spin Ice: Toward Magnonic Functional Materials. \emph{APL Mater.}
  \textbf{2020}, \emph{8}, 040911\relax
\mciteBstWouldAddEndPuncttrue
\mciteSetBstMidEndSepPunct{\mcitedefaultmidpunct}
{\mcitedefaultendpunct}{\mcitedefaultseppunct}\relax
\EndOfBibitem
\bibitem[Locatelli \latin{et~al.}(2014)Locatelli, Cros, and
  Grollier]{locatelli2014spin}
Locatelli,~N.; Cros,~V.; Grollier,~J. Spin-Torque Building Blocks. \emph{Nat.
  Mater.} \textbf{2014}, \emph{13}, 11--20\relax
\mciteBstWouldAddEndPuncttrue
\mciteSetBstMidEndSepPunct{\mcitedefaultmidpunct}
{\mcitedefaultendpunct}{\mcitedefaultseppunct}\relax
\EndOfBibitem
\bibitem[Lenk \latin{et~al.}(2011)Lenk, Ulrichs, Garbs, and
  M{\"u}nzenberg]{lenk2011building}
Lenk,~B.; Ulrichs,~H.; Garbs,~F.; M{\"u}nzenberg,~M. The Building Blocks of
  Magnonics. \emph{Phys. Rep.} \textbf{2011}, \emph{507}, 107--136\relax
\mciteBstWouldAddEndPuncttrue
\mciteSetBstMidEndSepPunct{\mcitedefaultmidpunct}
{\mcitedefaultendpunct}{\mcitedefaultseppunct}\relax
\EndOfBibitem
\bibitem[Pirro \latin{et~al.}(2014)Pirro, Br{\"a}cher, Chumak, L{\"a}gel, Dubs,
  Surzhenko, G{\"o}rnert, Leven, and Hillebrands]{pirro2014spin}
Pirro,~P.; Br{\"a}cher,~T.; Chumak,~A.~V.; L{\"a}gel,~B.; Dubs,~C.;
  Surzhenko,~O.; G{\"o}rnert,~P.; Leven,~B.; Hillebrands,~B. Spin-Wave
  Excitation and Propagation in Microstructured Waveguides of Yttrium Iron
  Garnet/Pt Bilayers. \emph{Appl. Phys. Lett.} \textbf{2014}, \emph{104},
  012402\relax
\mciteBstWouldAddEndPuncttrue
\mciteSetBstMidEndSepPunct{\mcitedefaultmidpunct}
{\mcitedefaultendpunct}{\mcitedefaultseppunct}\relax
\EndOfBibitem
\bibitem[Garcia-Sanchez \latin{et~al.}(2015)Garcia-Sanchez, Borys, Soucaille,
  Adam, Stamps, and Kim]{garcia2015narrow}
Garcia-Sanchez,~F.; Borys,~P.; Soucaille,~R.; Adam,~J.-P.; Stamps,~R.~L.;
  Kim,~J.-V. Narrow Magnonic Waveguides Based on Domain Walls. \emph{Phys. Rev.
  Lett.} \textbf{2015}, \emph{114}, 247206\relax
\mciteBstWouldAddEndPuncttrue
\mciteSetBstMidEndSepPunct{\mcitedefaultmidpunct}
{\mcitedefaultendpunct}{\mcitedefaultseppunct}\relax
\EndOfBibitem
\bibitem[Vogt \latin{et~al.}(2014)Vogt, Fradin, Pearson, Sebastian, Bader,
  Hillebrands, Hoffmann, and Schultheiss]{vogt2014realization}
Vogt,~K.; Fradin,~F.~Y.; Pearson,~J.~E.; Sebastian,~T.; Bader,~S.~D.;
  Hillebrands,~B.; Hoffmann,~A.; Schultheiss,~H. Realization of a Spin-Wave
  Multiplexer. \emph{Nat. Commun.} \textbf{2014}, \emph{5}, 1--5\relax
\mciteBstWouldAddEndPuncttrue
\mciteSetBstMidEndSepPunct{\mcitedefaultmidpunct}
{\mcitedefaultendpunct}{\mcitedefaultseppunct}\relax
\EndOfBibitem
\bibitem[Heussner \latin{et~al.}(2018)Heussner, Nabinger, Fischer, Br{\"a}cher,
  Serga, Hillebrands, and Pirro]{heussner2018frequency}
Heussner,~F.; Nabinger,~M.; Fischer,~T.; Br{\"a}cher,~T.; Serga,~A.~A.;
  Hillebrands,~B.; Pirro,~P. Frequency-Division Multiplexing in Magnonic Logic
  Networks Based on Caustic-Like Spin-Wave Beams. \emph{Phys. Status Solidi
  RRL} \textbf{2018}, \emph{12}, 1800409\relax
\mciteBstWouldAddEndPuncttrue
\mciteSetBstMidEndSepPunct{\mcitedefaultmidpunct}
{\mcitedefaultendpunct}{\mcitedefaultseppunct}\relax
\EndOfBibitem
\bibitem[Davies \latin{et~al.}(2015)Davies, Sadovnikov, Grishin, Sharaevsky,
  Nikitov, and Kruglyak]{davies2015field}
Davies,~C.~S.; Sadovnikov,~A.~V.; Grishin,~S.~V.; Sharaevsky,~Y.~P.;
  Nikitov,~S.~A.; Kruglyak,~V.~V. Field-Controlled Phase-Rectified Magnonic
  Multiplexer. \emph{IEEE Trans. Magn.} \textbf{2015}, \emph{51}, 1--4\relax
\mciteBstWouldAddEndPuncttrue
\mciteSetBstMidEndSepPunct{\mcitedefaultmidpunct}
{\mcitedefaultendpunct}{\mcitedefaultseppunct}\relax
\EndOfBibitem
\bibitem[Hansen \latin{et~al.}(2009)Hansen, Demidov, and
  Demokritov]{hansen2009dual}
Hansen,~U.-H.; Demidov,~V.~E.; Demokritov,~S.~O. Dual-Function Phase Shifter
  for Spin-Wave Logic Applications. \emph{Appl. Phys. Lett.} \textbf{2009},
  \emph{94}, 252502\relax
\mciteBstWouldAddEndPuncttrue
\mciteSetBstMidEndSepPunct{\mcitedefaultmidpunct}
{\mcitedefaultendpunct}{\mcitedefaultseppunct}\relax
\EndOfBibitem
\bibitem[Ustinov \latin{et~al.}(2014)Ustinov, Kalinikos, and
  Srinivasan]{ustinov2014nonlinear}
Ustinov,~A.~B.; Kalinikos,~B.~A.; Srinivasan,~G. Nonlinear Microwave Phase
  Shifter on Electromagnetic-Spin Waves. \emph{Tech. Phys.} \textbf{2014},
  \emph{59}, 1406--1408\relax
\mciteBstWouldAddEndPuncttrue
\mciteSetBstMidEndSepPunct{\mcitedefaultmidpunct}
{\mcitedefaultendpunct}{\mcitedefaultseppunct}\relax
\EndOfBibitem
\bibitem[Kostylev \latin{et~al.}(2007)Kostylev, Serga, Schneider, Neumann,
  Leven, Hillebrands, and Stamps]{kostylev2007resonant}
Kostylev,~M.~P.; Serga,~A.~A.; Schneider,~T.; Neumann,~T.; Leven,~B.;
  Hillebrands,~B.; Stamps,~R.~L. Resonant and Nonresonant Scattering of
  Dipole-Dominated Spin Waves from a Region of Inhomogeneous Magnetic Field in
  a Ferromagnetic Film. \emph{Phys. Rev. B} \textbf{2007}, \emph{76},
  184419\relax
\mciteBstWouldAddEndPuncttrue
\mciteSetBstMidEndSepPunct{\mcitedefaultmidpunct}
{\mcitedefaultendpunct}{\mcitedefaultseppunct}\relax
\EndOfBibitem
\bibitem[Fetisov and Patton(1999)Fetisov, and Patton]{fetisov1999microwave}
Fetisov,~Y.~K.; Patton,~C.~E. Microwave Bistability in a Magnetostatic Wave
  Interferometer with External Feedback. \emph{IEEE Trans. Magn.}
  \textbf{1999}, \emph{35}, 1024--1036\relax
\mciteBstWouldAddEndPuncttrue
\mciteSetBstMidEndSepPunct{\mcitedefaultmidpunct}
{\mcitedefaultendpunct}{\mcitedefaultseppunct}\relax
\EndOfBibitem
\bibitem[Schneider \latin{et~al.}(2008)Schneider, Serga, Leven, Hillebrands,
  Stamps, and Kostylev]{schneider2008realization}
Schneider,~T.; Serga,~A.~A.; Leven,~B.; Hillebrands,~B.; Stamps,~R.~L.;
  Kostylev,~M.~P. Realization of Spin-Wave Logic Gates. \emph{Appl. Phys.
  Lett.} \textbf{2008}, \emph{92}, 022505\relax
\mciteBstWouldAddEndPuncttrue
\mciteSetBstMidEndSepPunct{\mcitedefaultmidpunct}
{\mcitedefaultendpunct}{\mcitedefaultseppunct}\relax
\EndOfBibitem
\bibitem[Klingler \latin{et~al.}(2015)Klingler, Pirro, Br{\"a}cher, Leven,
  Hillebrands, and Chumak]{klingler2015spin}
Klingler,~S.; Pirro,~P.; Br{\"a}cher,~T.; Leven,~B.; Hillebrands,~B.;
  Chumak,~A.~V. Spin-Wave Logic Devices Based on Isotropic Forward Volume
  Magnetostatic Waves. \emph{Appl. Phys. Lett.} \textbf{2015}, \emph{106},
  212406\relax
\mciteBstWouldAddEndPuncttrue
\mciteSetBstMidEndSepPunct{\mcitedefaultmidpunct}
{\mcitedefaultendpunct}{\mcitedefaultseppunct}\relax
\EndOfBibitem
\bibitem[Ding \latin{et~al.}(2012)Ding, Kostylev, and
  Adeyeye]{ding2012realization}
Ding,~J.; Kostylev,~M.; Adeyeye,~A.~O. Realization of a Mesoscopic
  Reprogrammable Magnetic Logic Based on a Nanoscale Reconfigurable Magnonic
  Crystal. \emph{Appl. Phys. Lett.} \textbf{2012}, \emph{100}, 073114\relax
\mciteBstWouldAddEndPuncttrue
\mciteSetBstMidEndSepPunct{\mcitedefaultmidpunct}
{\mcitedefaultendpunct}{\mcitedefaultseppunct}\relax
\EndOfBibitem
\bibitem[Hertel \latin{et~al.}(2004)Hertel, Wulfhekel, and
  Kirschner]{hertel2004domain}
Hertel,~R.; Wulfhekel,~W.; Kirschner,~J. Domain-Wall Induced Phase Shifts in
  Spin Waves. \emph{Phys. Rev. Lett.} \textbf{2004}, \emph{93}, 257202\relax
\mciteBstWouldAddEndPuncttrue
\mciteSetBstMidEndSepPunct{\mcitedefaultmidpunct}
{\mcitedefaultendpunct}{\mcitedefaultseppunct}\relax
\EndOfBibitem
\bibitem[Dion \latin{et~al.}(2019)Dion, Arroo, Yamanoi, Kimura, Gartside,
  Cohen, Kurebayashi, and Branford]{dion2019tunable}
Dion,~T.; Arroo,~D.~M.; Yamanoi,~K.; Kimura,~T.; Gartside,~J.~C.; Cohen,~L.~F.;
  Kurebayashi,~H.; Branford,~W. Tunable Magnetization Dynamics in Artificial
  Spin Ice \textit{via} Shape Anisotropy Modification. \emph{Phys. Rev. B}
  \textbf{2019}, \emph{100}, 054433\relax
\mciteBstWouldAddEndPuncttrue
\mciteSetBstMidEndSepPunct{\mcitedefaultmidpunct}
{\mcitedefaultendpunct}{\mcitedefaultseppunct}\relax
\EndOfBibitem
\bibitem[Iacocca \latin{et~al.}(2020)Iacocca, Gliga, and
  Heinonen]{iacocca2020tailoring}
Iacocca,~E.; Gliga,~S.; Heinonen,~O.~G. Tailoring Spin-Wave Channels in a
  Reconfigurable Artificial Spin Ice. \emph{Phys. Rev. Appl.} \textbf{2020},
  \emph{13}, 044047\relax
\mciteBstWouldAddEndPuncttrue
\mciteSetBstMidEndSepPunct{\mcitedefaultmidpunct}
{\mcitedefaultendpunct}{\mcitedefaultseppunct}\relax
\EndOfBibitem
\bibitem[Sadovnikov \latin{et~al.}(2018)Sadovnikov, Gubanov, Sheshukova,
  Sharaevskii, and Nikitov]{sadovnikov2018spin}
Sadovnikov,~A.~V.; Gubanov,~V.~A.; Sheshukova,~S.~E.; Sharaevskii,~Y.~P.;
  Nikitov,~S.~A. Spin-Wave Drop Filter Based on Asymmetric Side-Coupled
  Magnonic Crystals. \emph{Phys. Rev. Appl.} \textbf{2018}, \emph{9},
  051002\relax
\mciteBstWouldAddEndPuncttrue
\mciteSetBstMidEndSepPunct{\mcitedefaultmidpunct}
{\mcitedefaultendpunct}{\mcitedefaultseppunct}\relax
\EndOfBibitem
\bibitem[Wang \latin{et~al.}(2015)Wang, Zeng, Lei, and Bi]{wang2015tunable}
Wang,~Q.; Zeng,~L.; Lei,~M.; Bi,~K. Tunable Metamaterial Bandstop Filter Based
  on Ferromagnetic Resonance. \emph{AIP Adv.} \textbf{2015}, \emph{5},
  077145\relax
\mciteBstWouldAddEndPuncttrue
\mciteSetBstMidEndSepPunct{\mcitedefaultmidpunct}
{\mcitedefaultendpunct}{\mcitedefaultseppunct}\relax
\EndOfBibitem
\bibitem[Semenova and Berkov(2013)Semenova, and Berkov]{semenova2013spin}
Semenova,~E.~K.; Berkov,~D.~V. Spin Wave Propagation Through an Antidot Lattice
  and a Concept of a Tunable Magnonic Filter. \emph{J. Appl. Phys.}
  \textbf{2013}, \emph{114}, 013905\relax
\mciteBstWouldAddEndPuncttrue
\mciteSetBstMidEndSepPunct{\mcitedefaultmidpunct}
{\mcitedefaultendpunct}{\mcitedefaultseppunct}\relax
\EndOfBibitem
\bibitem[Ma \latin{et~al.}(2011)Ma, Lim, Wang, Piramanayagam, Ng, and
  Kuok]{ma2011micromagnetic}
Ma,~F.~S.; Lim,~H.~S.; Wang,~Z.~K.; Piramanayagam,~S.~N.; Ng,~S.~C.;
  Kuok,~M.~H. Micromagnetic Study of Spin Wave Propagation in Bicomponent
  Magnonic Crystal Waveguides. \emph{Appl. Phys. Lett.} \textbf{2011},
  \emph{98}, 153107\relax
\mciteBstWouldAddEndPuncttrue
\mciteSetBstMidEndSepPunct{\mcitedefaultmidpunct}
{\mcitedefaultendpunct}{\mcitedefaultseppunct}\relax
\EndOfBibitem
\bibitem[Kim \latin{et~al.}(2009)Kim, Lee, and Han]{kim2009gigahertz}
Kim,~S.-K.; Lee,~K.-S.; Han,~D.-S. A Gigahertz-Range Spin-Wave Filter Composed
  of Width-Modulated Nanostrip Magnonic-Crystal Waveguides. \emph{Appl. Phys.
  Lett.} \textbf{2009}, \emph{95}, 082507\relax
\mciteBstWouldAddEndPuncttrue
\mciteSetBstMidEndSepPunct{\mcitedefaultmidpunct}
{\mcitedefaultendpunct}{\mcitedefaultseppunct}\relax
\EndOfBibitem
\bibitem[Louis \latin{et~al.}(2016)Louis, Lisenkov, Nikitov, Tyberkevych, and
  Slavin]{louis2016bias}
Louis,~S.; Lisenkov,~I.; Nikitov,~S.; Tyberkevych,~V.; Slavin,~A. Bias-Free
  Spin-Wave Phase Shifter for Magnonic Logic. \emph{AIP Adv.} \textbf{2016},
  \emph{6}, 065103\relax
\mciteBstWouldAddEndPuncttrue
\mciteSetBstMidEndSepPunct{\mcitedefaultmidpunct}
{\mcitedefaultendpunct}{\mcitedefaultseppunct}\relax
\EndOfBibitem
\bibitem[Vogt \latin{et~al.}(2012)Vogt, Schultheiss, Jain, Pearson, Hoffmann,
  Bader, and Hillebrands]{vogt2012spin}
Vogt,~K.; Schultheiss,~H.; Jain,~S.; Pearson,~J.~E.; Hoffmann,~A.;
  Bader,~S.~D.; Hillebrands,~B. Spin Waves Turning a Corner. \emph{Appl. Phys.
  Lett.} \textbf{2012}, \emph{101}, 042410\relax
\mciteBstWouldAddEndPuncttrue
\mciteSetBstMidEndSepPunct{\mcitedefaultmidpunct}
{\mcitedefaultendpunct}{\mcitedefaultseppunct}\relax
\EndOfBibitem
\bibitem[Davies \latin{et~al.}(2015)Davies, Francis, Sadovnikov, Chertopalov,
  Bryan, Grishin, Allwood, Sharaevskii, Nikitov, and
  Kruglyak]{davies2015towards}
Davies,~C.~S.; Francis,~A.; Sadovnikov,~A.~V.; Chertopalov,~S.~V.;
  Bryan,~M.~T.; Grishin,~S.~V.; Allwood,~D.~A.; Sharaevskii,~Y.~P.;
  Nikitov,~S.~A.; Kruglyak,~V.~V. Towards Graded-Index Magnonics: Steering Spin
  Waves in Magnonic Networks. \emph{Phys. Rev. B} \textbf{2015}, \emph{92},
  020408\relax
\mciteBstWouldAddEndPuncttrue
\mciteSetBstMidEndSepPunct{\mcitedefaultmidpunct}
{\mcitedefaultendpunct}{\mcitedefaultseppunct}\relax
\EndOfBibitem
\bibitem[Albisetti \latin{et~al.}(2018)Albisetti, Petti, Sala, Silvani, Tacchi,
  Finizio, Wintz, Cal{\`o}, Zheng, Raabe, Riedo, and
  Bertacco]{albisetti2018nanoscale}
Albisetti,~E.; Petti,~D.; Sala,~G.; Silvani,~R.; Tacchi,~S.; Finizio,~S.;
  Wintz,~S.; Cal{\`o},~A.; Zheng,~X.; Raabe,~J.; Riedo,~E.; Bertacco,~R.
  Nanoscale Spin-Wave Circuits Based on Engineered Reconfigurable
  Spin-Textures. \emph{Commun. Phys.} \textbf{2018}, \emph{1}, 1--8\relax
\mciteBstWouldAddEndPuncttrue
\mciteSetBstMidEndSepPunct{\mcitedefaultmidpunct}
{\mcitedefaultendpunct}{\mcitedefaultseppunct}\relax
\EndOfBibitem
\bibitem[Wang \latin{et~al.}(2016)Wang, Xiao, Snezhko, Xu, Ocola, Divan,
  Pearson, Crabtree, and Kwok]{wang2016rewritable}
Wang,~Y.-L.; Xiao,~Z.-L.; Snezhko,~A.; Xu,~J.; Ocola,~L.~E.; Divan,~R.;
  Pearson,~J.~E.; Crabtree,~G.~W.; Kwok,~W.-K. Rewritable Artificial Magnetic
  Charge Ice. \emph{Science} \textbf{2016}, \emph{352}, 962--966\relax
\mciteBstWouldAddEndPuncttrue
\mciteSetBstMidEndSepPunct{\mcitedefaultmidpunct}
{\mcitedefaultendpunct}{\mcitedefaultseppunct}\relax
\EndOfBibitem
\bibitem[Gartside \latin{et~al.}(2016)Gartside, Burn, Cohen, and
  Branford]{gartside2016novel}
Gartside,~J.~C.; Burn,~D.~M.; Cohen,~L.~F.; Branford,~W.~R. A Novel Method for
  the Injection and Manipulation of Magnetic Charge States in Nanostructures.
  \emph{Sci. Rep.} \textbf{2016}, \emph{6}, 32864\relax
\mciteBstWouldAddEndPuncttrue
\mciteSetBstMidEndSepPunct{\mcitedefaultmidpunct}
{\mcitedefaultendpunct}{\mcitedefaultseppunct}\relax
\EndOfBibitem
\bibitem[Gartside \latin{et~al.}(2018)Gartside, Arroo, Burn, Bemmer,
  Moskalenko, Cohen, and Branford]{gartside2018realization}
Gartside,~J.~C.; Arroo,~D.~M.; Burn,~D.~M.; Bemmer,~V.~L.; Moskalenko,~A.;
  Cohen,~L.~F.; Branford,~W.~R. Realization of Ground State in Artificial
  Kagome Spin Ice \textit{via} Topological Defect-Driven Magnetic Writing.
  \emph{Nat. Nanotechnol.} \textbf{2018}, \emph{13}, 53--58\relax
\mciteBstWouldAddEndPuncttrue
\mciteSetBstMidEndSepPunct{\mcitedefaultmidpunct}
{\mcitedefaultendpunct}{\mcitedefaultseppunct}\relax
\EndOfBibitem
\bibitem[Gartside \latin{et~al.}(2020)Gartside, Jung, Yoo, Arroo, Vanstone,
  Dion, Stenning, and Branford]{gartside2020current}
Gartside,~J.~C.; Jung,~S.~G.; Yoo,~S.~Y.; Arroo,~D.~M.; Vanstone,~A.; Dion,~T.;
  Stenning,~K.~D.; Branford,~W.~R. Current-Controlled Nanomagnetic Writing for
  Reconfigurable Magnonic Crystals. \emph{Commun. Phys.} \textbf{2020},
  \emph{3}, 219\relax
\mciteBstWouldAddEndPuncttrue
\mciteSetBstMidEndSepPunct{\mcitedefaultmidpunct}
{\mcitedefaultendpunct}{\mcitedefaultseppunct}\relax
\EndOfBibitem
\bibitem[Cowburn \latin{et~al.}(1999)Cowburn, Koltsov, Adeyeye, Welland, and
  Tricker]{cowburn1999single}
Cowburn,~R.~P.; Koltsov,~D.~K.; Adeyeye,~A.~O.; Welland,~M.~E.; Tricker,~D.~M.
  Single-Domain Circular Nanomagnets. \emph{Phys. Rev. Lett.} \textbf{1999},
  \emph{83}, 1042\relax
\mciteBstWouldAddEndPuncttrue
\mciteSetBstMidEndSepPunct{\mcitedefaultmidpunct}
{\mcitedefaultendpunct}{\mcitedefaultseppunct}\relax
\EndOfBibitem
\bibitem[Jubert and Allenspach(2004)Jubert, and
  Allenspach]{jubert2004analytical}
Jubert,~P.~O.; Allenspach,~R. Analytical Approach to the
  Single-Domain-to-Vortex Transition in Small Magnetic Disks. \emph{Phys. Rev.
  B} \textbf{2004}, \emph{70}, 144402\relax
\mciteBstWouldAddEndPuncttrue
\mciteSetBstMidEndSepPunct{\mcitedefaultmidpunct}
{\mcitedefaultendpunct}{\mcitedefaultseppunct}\relax
\EndOfBibitem
\bibitem[Chung \latin{et~al.}(2010)Chung, McMichael, Pierce, and
  Unguris]{chung2010phase}
Chung,~S.-H.; McMichael,~R.~D.; Pierce,~D.~T.; Unguris,~J. Phase Diagram of
  Magnetic Nanodisks Measured by Scanning Electron Microscopy with Polarization
  Analysis. \emph{Phys. Rev. B} \textbf{2010}, \emph{81}, 024410\relax
\mciteBstWouldAddEndPuncttrue
\mciteSetBstMidEndSepPunct{\mcitedefaultmidpunct}
{\mcitedefaultendpunct}{\mcitedefaultseppunct}\relax
\EndOfBibitem
\bibitem[Metlov and Lee(2008)Metlov, and Lee]{metlov2008map}
Metlov,~K.~L.; Lee,~Y.~P. Map of Metastable States for Thin Circular Magnetic
  Nanocylinders. \emph{Appl. Phys. Lett.} \textbf{2008}, \emph{92},
  112506\relax
\mciteBstWouldAddEndPuncttrue
\mciteSetBstMidEndSepPunct{\mcitedefaultmidpunct}
{\mcitedefaultendpunct}{\mcitedefaultseppunct}\relax
\EndOfBibitem
\bibitem[{\"O}stman \latin{et~al.}(2014){\"O}stman, Arnalds, Melander,
  Kapaklis, P{\'a}lsson, Saw, Verschuuren, Kronast, Papaioannou, Fadley, and
  Hj{\"o}rvarsson]{ostman2014hysteresis}
{\"O}stman,~E.; Arnalds,~U.~B.; Melander,~E.; Kapaklis,~V.; P{\'a}lsson,~G.~K.;
  Saw,~A.~Y.; Verschuuren,~M.~A.; Kronast,~F.; Papaioannou,~E.~T.;
  Fadley,~C.~S.; Hj{\"o}rvarsson,~B. Hysteresis-Free Switching between Vortex
  and Collinear Magnetic States. \emph{New J. Phys.} \textbf{2014}, \emph{16},
  053002\relax
\mciteBstWouldAddEndPuncttrue
\mciteSetBstMidEndSepPunct{\mcitedefaultmidpunct}
{\mcitedefaultendpunct}{\mcitedefaultseppunct}\relax
\EndOfBibitem
\bibitem[G{\'e}lvez and Pati{\~n}o(2019)G{\'e}lvez, and
  Pati{\~n}o]{gelvez2019coercive}
G{\'e}lvez,~C.~F.; Pati{\~n}o,~E.~J. Coercive Field Enhancement in Co
  Nanodisks: Single-Domain to Vortex Switching. \emph{J. Phys.: Condens.
  Matter} \textbf{2019}, \emph{31}, 13LT01\relax
\mciteBstWouldAddEndPuncttrue
\mciteSetBstMidEndSepPunct{\mcitedefaultmidpunct}
{\mcitedefaultendpunct}{\mcitedefaultseppunct}\relax
\EndOfBibitem
\bibitem[Van~Waeyenberge \latin{et~al.}(2006)Van~Waeyenberge, Puzic, Stoll,
  Chou, Tyliszczak, Hertel, F{\"a}hnle, Br{\"u}ckl, Rott, Reiss, Neudecker,
  Weiss, Back, and Sch{\"u}tz]{van2006magnetic}
Van~Waeyenberge,~B.; Puzic,~A.; Stoll,~H.; Chou,~K.; Tyliszczak,~T.;
  Hertel,~R.; F{\"a}hnle,~M.; Br{\"u}ckl,~H.; Rott,~K.; Reiss,~G.;
  Neudecker,~I.; Weiss,~D.; Back,~C.~H.; Sch{\"u}tz,~G. Magnetic Vortex Core
  Reversal by Excitation with Short Bursts of an Alternating Field.
  \emph{Nature} \textbf{2006}, \emph{444}, 461--464\relax
\mciteBstWouldAddEndPuncttrue
\mciteSetBstMidEndSepPunct{\mcitedefaultmidpunct}
{\mcitedefaultendpunct}{\mcitedefaultseppunct}\relax
\EndOfBibitem
\bibitem[Yamada \latin{et~al.}(2007)Yamada, Kasai, Nakatani, Kobayashi, Kohno,
  Thiaville, and Ono]{yamada2007electrical}
Yamada,~K.; Kasai,~S.; Nakatani,~Y.; Kobayashi,~K.; Kohno,~H.; Thiaville,~A.;
  Ono,~T. Electrical Switching of the Vortex Core in a Magnetic Disk.
  \emph{Nat. Mater.} \textbf{2007}, \emph{6}, 270--273\relax
\mciteBstWouldAddEndPuncttrue
\mciteSetBstMidEndSepPunct{\mcitedefaultmidpunct}
{\mcitedefaultendpunct}{\mcitedefaultseppunct}\relax
\EndOfBibitem
\bibitem[Uhl{\'\i}{\v{r}} \latin{et~al.}(2013)Uhl{\'\i}{\v{r}}, Urb{\'a}nek,
  Hlad{\'\i}k, Spousta, Im, Fischer, Eibagi, Kan, Fullerton, and
  {\v{S}}ikola]{uhlivr2013dynamic}
Uhl{\'\i}{\v{r}},~V.; Urb{\'a}nek,~M.; Hlad{\'\i}k,~L.; Spousta,~J.; Im,~M.~Y.;
  Fischer,~P.; Eibagi,~N.; Kan,~J.~J.; Fullerton,~E.~E.; {\v{S}}ikola,~T.
  Dynamic Switching of the Spin Circulation in Tapered Magnetic Nanodisks.
  \emph{Nat. Nanotechnol.} \textbf{2013}, \emph{8}, 341\relax
\mciteBstWouldAddEndPuncttrue
\mciteSetBstMidEndSepPunct{\mcitedefaultmidpunct}
{\mcitedefaultendpunct}{\mcitedefaultseppunct}\relax
\EndOfBibitem
\bibitem[Agramunt-Puig \latin{et~al.}(2014)Agramunt-Puig, Del-Valle, Navau, and
  Sanchez]{agramunt2014controlling}
Agramunt-Puig,~S.; Del-Valle,~N.; Navau,~C.; Sanchez,~A. Controlling Vortex
  Chirality and Polarity by Geometry in Magnetic Nanodots. \emph{Appl. Phys.
  Lett.} \textbf{2014}, \emph{104}, 012407\relax
\mciteBstWouldAddEndPuncttrue
\mciteSetBstMidEndSepPunct{\mcitedefaultmidpunct}
{\mcitedefaultendpunct}{\mcitedefaultseppunct}\relax
\EndOfBibitem
\bibitem[Dumas \latin{et~al.}(2011)Dumas, Gilbert, Eibagi, and
  Liu]{dumas2011chirality}
Dumas,~R.~K.; Gilbert,~D.~A.; Eibagi,~N.; Liu,~K. Chirality Control
  \textit{via} Double Vortices in Asymmetric Co Dots. \emph{Phys. Rev. B}
  \textbf{2011}, \emph{83}, 060415\relax
\mciteBstWouldAddEndPuncttrue
\mciteSetBstMidEndSepPunct{\mcitedefaultmidpunct}
{\mcitedefaultendpunct}{\mcitedefaultseppunct}\relax
\EndOfBibitem
\bibitem[Haldar and Adeyeye(2015)Haldar, and Adeyeye]{haldar2015vortex}
Haldar,~A.; Adeyeye,~A.~O. Vortex Chirality Control in Circular Disks Using
  Dipole-Coupled Nanomagnets. \emph{Appl. Phys. Lett.} \textbf{2015},
  \emph{106}, 032404\relax
\mciteBstWouldAddEndPuncttrue
\mciteSetBstMidEndSepPunct{\mcitedefaultmidpunct}
{\mcitedefaultendpunct}{\mcitedefaultseppunct}\relax
\EndOfBibitem
\bibitem[Tacchi \latin{et~al.}(2011)Tacchi, Montoncello, Madami, Gubbiotti,
  Carlotti, Giovannini, Zivieri, Nizzoli, Jain, Adeyeye, and
  Singh]{PhysRevLett.107.127204}
Tacchi,~S.; Montoncello,~F.; Madami,~M.; Gubbiotti,~G.; Carlotti,~G.;
  Giovannini,~L.; Zivieri,~R.; Nizzoli,~F.; Jain,~S.; Adeyeye,~A.~O.; Singh,~N.
  Band Diagram of Spin Waves in a Two-Dimensional Magnonic Crystal. \emph{Phys.
  Rev. Lett.} \textbf{2011}, \emph{107}, 127204\relax
\mciteBstWouldAddEndPuncttrue
\mciteSetBstMidEndSepPunct{\mcitedefaultmidpunct}
{\mcitedefaultendpunct}{\mcitedefaultseppunct}\relax
\EndOfBibitem
\bibitem[Huber and Grundler(2011)Huber, and Grundler]{huber2011ferromagnetic}
Huber,~R.; Grundler,~D. Ferromagnetic Nanodisks for Magnonic Crystals and
  Waveguides. Spintronics IV. 2011; p 81000D\relax
\mciteBstWouldAddEndPuncttrue
\mciteSetBstMidEndSepPunct{\mcitedefaultmidpunct}
{\mcitedefaultendpunct}{\mcitedefaultseppunct}\relax
\EndOfBibitem
\bibitem[Kaffash \latin{et~al.}(2020)Kaffash, Bang, Lendinez, Hoffmann,
  Ketterson, and Jungfleisch]{kaffash2020control}
Kaffash,~M.~T.; Bang,~W.; Lendinez,~S.; Hoffmann,~A.; Ketterson,~J.~B.;
  Jungfleisch,~M.~B. Control of Spin Dynamics in Artificial Honeycomb
  Spin-Ice-Based Nanodisks. \emph{Phys. Rev. B} \textbf{2020}, \emph{101},
  174424\relax
\mciteBstWouldAddEndPuncttrue
\mciteSetBstMidEndSepPunct{\mcitedefaultmidpunct}
{\mcitedefaultendpunct}{\mcitedefaultseppunct}\relax
\EndOfBibitem
\bibitem[Kumar \latin{et~al.}(2014)Kumar, Barman, and
  Barman]{kumar2014magnetic}
Kumar,~D.; Barman,~S.; Barman,~A. Magnetic Vortex Based Transistor Operations.
  \emph{Sci. Rep.} \textbf{2014}, \emph{4}, 4108\relax
\mciteBstWouldAddEndPuncttrue
\mciteSetBstMidEndSepPunct{\mcitedefaultmidpunct}
{\mcitedefaultendpunct}{\mcitedefaultseppunct}\relax
\EndOfBibitem
\bibitem[Shibata \latin{et~al.}(2003)Shibata, Shigeto, and
  Otani]{shibata2003dynamics}
Shibata,~J.; Shigeto,~K.; Otani,~Y. Dynamics of Magnetostatically Coupled
  Vortices in Magnetic Nanodisks. \emph{Phys. Rev. B} \textbf{2003}, \emph{67},
  224404\relax
\mciteBstWouldAddEndPuncttrue
\mciteSetBstMidEndSepPunct{\mcitedefaultmidpunct}
{\mcitedefaultendpunct}{\mcitedefaultseppunct}\relax
\EndOfBibitem
\bibitem[Sugimoto \latin{et~al.}(2011)Sugimoto, Fukuma, Kasai, Kimura, Barman,
  and Otani]{sugimoto2011dynamics}
Sugimoto,~S.; Fukuma,~Y.; Kasai,~S.; Kimura,~T.; Barman,~A.; Otani,~Y. Dynamics
  of Coupled Vortices in a Pair of Ferromagnetic Disks. \emph{Phys. Rev. Lett.}
  \textbf{2011}, \emph{106}, 197203\relax
\mciteBstWouldAddEndPuncttrue
\mciteSetBstMidEndSepPunct{\mcitedefaultmidpunct}
{\mcitedefaultendpunct}{\mcitedefaultseppunct}\relax
\EndOfBibitem
\bibitem[Vogel \latin{et~al.}(2011)Vogel, Kamionka, Martens, Drews, Chou,
  Tyliszczak, Stoll, Van~Waeyenberge, and Meier]{vogel2011coupled}
Vogel,~A.; Kamionka,~T.; Martens,~M.; Drews,~A.; Chou,~K.~W.; Tyliszczak,~T.;
  Stoll,~H.; Van~Waeyenberge,~B.; Meier,~G. Coupled Vortex Oscillations in
  Spatially Separated Permalloy Squares. \emph{Phys. Rev. Lett.} \textbf{2011},
  \emph{106}, 137201\relax
\mciteBstWouldAddEndPuncttrue
\mciteSetBstMidEndSepPunct{\mcitedefaultmidpunct}
{\mcitedefaultendpunct}{\mcitedefaultseppunct}\relax
\EndOfBibitem
\bibitem[Jung \latin{et~al.}(2011)Jung, Lee, Jeong, Choi, Yu, Han, Vogel,
  Bocklage, Meier, Im, Fischer, and Kim]{jung2011tunable}
Jung,~H.; Lee,~K.-S.; Jeong,~D.-E.; Choi,~Y.-S.; Yu,~Y.-S.; Han,~D.-S.;
  Vogel,~A.; Bocklage,~L.; Meier,~G.; Im,~M.-Y.; Fischer,~P.; Kim,~S.-K.
  Tunable Negligible-Loss Energy Transfer between Dipolar-Coupled Magnetic
  Disks by Stimulated Vortex Gyration. \emph{Sci. Rep.} \textbf{2011},
  \emph{1}, 59\relax
\mciteBstWouldAddEndPuncttrue
\mciteSetBstMidEndSepPunct{\mcitedefaultmidpunct}
{\mcitedefaultendpunct}{\mcitedefaultseppunct}\relax
\EndOfBibitem
\bibitem[Barman \latin{et~al.}(2016)Barman, Saha, Mondal, Kumar, and
  Barman]{barman2016enhanced}
Barman,~S.; Saha,~S.; Mondal,~S.; Kumar,~D.; Barman,~A. Enhanced Amplification
  and Fan-Out Operation in an All-Magnetic Transistor. \emph{Sci. Rep.}
  \textbf{2016}, \emph{6}, 33360\relax
\mciteBstWouldAddEndPuncttrue
\mciteSetBstMidEndSepPunct{\mcitedefaultmidpunct}
{\mcitedefaultendpunct}{\mcitedefaultseppunct}\relax
\EndOfBibitem
\bibitem[Mondal \latin{et~al.}(2020)Mondal, Barman, and
  Barman]{mondal2020magnetic}
Mondal,~S.; Barman,~S.; Barman,~A. Magnetic Vortex Transistor Based Tri-State
  Buffer Switch. \emph{J. Magn. Magn. Mater.} \textbf{2020}, \emph{502},
  166520\relax
\mciteBstWouldAddEndPuncttrue
\mciteSetBstMidEndSepPunct{\mcitedefaultmidpunct}
{\mcitedefaultendpunct}{\mcitedefaultseppunct}\relax
\EndOfBibitem
\bibitem[Mondal \latin{et~al.}(2020)Mondal, Banerjee, Adhikari, Chaurasiya,
  Choudhury, Sinha, Barman, and Barman]{mondal2020spin}
Mondal,~A.~K.; Banerjee,~C.; Adhikari,~A.; Chaurasiya,~A.~K.; Choudhury,~S.;
  Sinha,~J.; Barman,~S.; Barman,~A. Spin-Texture Driven Reconfigurable
  Magnonics in Chains of Connected Ni\textsubscript{80}Fe\textsubscript{20}
  Submicron Dots. \emph{Phys. Rev. B} \textbf{2020}, \emph{101}, 224426\relax
\mciteBstWouldAddEndPuncttrue
\mciteSetBstMidEndSepPunct{\mcitedefaultmidpunct}
{\mcitedefaultendpunct}{\mcitedefaultseppunct}\relax
\EndOfBibitem
\bibitem[Burgos-Parra \latin{et~al.}(2019)Burgos-Parra, Keatley, Sani,
  Durrenfeld, {\AA}kerman, and Hicken]{burgos2019time}
Burgos-Parra,~E.; Keatley,~P.~S.; Sani,~S.; Durrenfeld,~P.; {\AA}kerman,~J.;
  Hicken,~R.~J. Time-Resolved Imaging of Magnetization Dynamics in Double
  Nanocontact Spin Torque Vortex Oscillator Devices. \emph{Phys. Rev. B}
  \textbf{2019}, \emph{100}, 134439\relax
\mciteBstWouldAddEndPuncttrue
\mciteSetBstMidEndSepPunct{\mcitedefaultmidpunct}
{\mcitedefaultendpunct}{\mcitedefaultseppunct}\relax
\EndOfBibitem
\bibitem[Ramasubramanian \latin{et~al.}(2020)Ramasubramanian, K{\'a}kay,
  Fowley, Yildirim, Matthes, Sorokin, Titova, Hilliard, Boettger, H{\"u}bner,
  Gemming, Schulz, Kronast, Makarov, Fassbender, and
  Deac]{ramasubramanian2020tunable}
Ramasubramanian,~L.; K{\'a}kay,~A.; Fowley,~C.; Yildirim,~O.; Matthes,~P.;
  Sorokin,~S.; Titova,~A.; Hilliard,~D.; Boettger,~R.; H{\"u}bner,~R.;
  Gemming,~S.; Schulz,~S.~E.; Kronast,~F.; Makarov,~D.; Fassbender,~J.;
  Deac,~A. Tunable Magnetic Vortex Dynamics in Ion-Implanted Permalloy Disks.
  \emph{ACS Appl. Mater. Interfaces} \textbf{2020}, \emph{12},
  27812--27818\relax
\mciteBstWouldAddEndPuncttrue
\mciteSetBstMidEndSepPunct{\mcitedefaultmidpunct}
{\mcitedefaultendpunct}{\mcitedefaultseppunct}\relax
\EndOfBibitem
\bibitem[Morales \latin{et~al.}(2020)Morales, Flores, Vargas, Giuliani,
  Schuller, and Monton]{morales2020ultradense}
Morales,~R.; Flores,~A.~N.; Vargas,~N.~M.; Giuliani,~J.; Schuller,~I.~K.;
  Monton,~C. Ultradense Arrays of Sub-100 nm Co/CoO Nanodisks for Spintronics
  Applications. \emph{ACS Appl. Nano Mater.} \textbf{2020}, \emph{3},
  4037--4044\relax
\mciteBstWouldAddEndPuncttrue
\mciteSetBstMidEndSepPunct{\mcitedefaultmidpunct}
{\mcitedefaultendpunct}{\mcitedefaultseppunct}\relax
\EndOfBibitem
\bibitem[Li \latin{et~al.}(2020)Li, Cao, Wang, and Yan]{li2020second}
Li,~Z.-X.; Cao,~Y.; Wang,~X.~R.; Yan,~P. Second-Order Topological Solitonic
  Insulator in a Breathing Square Lattice of Magnetic Vortices. \emph{Phys.
  Rev. B} \textbf{2020}, \emph{101}, 184404\relax
\mciteBstWouldAddEndPuncttrue
\mciteSetBstMidEndSepPunct{\mcitedefaultmidpunct}
{\mcitedefaultendpunct}{\mcitedefaultseppunct}\relax
\EndOfBibitem
\bibitem[Cramer \latin{et~al.}(2018)Cramer, Fuhrmann, Ritzmann, Gall, Niizeki,
  Ramos, Qiu, Hou, Kikkawa, Sinova, Nowak, Saitoh, and
  Kl{\"a}ui]{cramer2018magnon}
Cramer,~J.; Fuhrmann,~F.; Ritzmann,~U.; Gall,~V.; Niizeki,~T.; Ramos,~R.;
  Qiu,~Z.; Hou,~D.; Kikkawa,~T.; Sinova,~J.; Nowak,~U.; Saitoh,~E.;
  Kl{\"a}ui,~M. Magnon Detection Using a Ferroic Collinear Multilayer Spin
  Valve. \emph{Nat. Commun.} \textbf{2018}, \emph{9}, 1--7\relax
\mciteBstWouldAddEndPuncttrue
\mciteSetBstMidEndSepPunct{\mcitedefaultmidpunct}
{\mcitedefaultendpunct}{\mcitedefaultseppunct}\relax
\EndOfBibitem
\bibitem[Wu \latin{et~al.}(2018)Wu, Huang, Fang, Yang, Wan, Yu, Feng, Wei, and
  Han]{wu2018magnon}
Wu,~H.; Huang,~L.; Fang,~C.; Yang,~B.~S.; Wan,~C.; Yu,~G.~Q.; Feng,~J.~F.;
  Wei,~H.~X.; Han,~X.~F. Magnon Valve Effect between Two Magnetic Insulators.
  \emph{Phys. Rev. Lett.} \textbf{2018}, \emph{120}, 097205\relax
\mciteBstWouldAddEndPuncttrue
\mciteSetBstMidEndSepPunct{\mcitedefaultmidpunct}
{\mcitedefaultendpunct}{\mcitedefaultseppunct}\relax
\EndOfBibitem
\bibitem[Cornelissen \latin{et~al.}(2018)Cornelissen, Liu, Van~Wees, and
  Duine]{cornelissen2018spin}
Cornelissen,~L.~J.; Liu,~J.; Van~Wees,~B.~J.; Duine,~R. Spin-Current-Controlled
  Modulation of the Magnon Spin Conductance in a Three-Terminal Magnon
  Transistor. \emph{Phys. Rev. Lett.} \textbf{2018}, \emph{120}, 097702\relax
\mciteBstWouldAddEndPuncttrue
\mciteSetBstMidEndSepPunct{\mcitedefaultmidpunct}
{\mcitedefaultendpunct}{\mcitedefaultseppunct}\relax
\EndOfBibitem
\bibitem[Magiera \latin{et~al.}(2012)Magiera, Hucht, Hinrichsen, Dahmen, and
  Wolf]{magiera2012magnetic}
Magiera,~M.~P.; Hucht,~A.; Hinrichsen,~H.; Dahmen,~S.~R.; Wolf,~D.~E. Magnetic
  Vortices Induced by a Moving Tip. \emph{Europhys. Lett.} \textbf{2012},
  \emph{100}, 27004\relax
\mciteBstWouldAddEndPuncttrue
\mciteSetBstMidEndSepPunct{\mcitedefaultmidpunct}
{\mcitedefaultendpunct}{\mcitedefaultseppunct}\relax
\EndOfBibitem
\bibitem[Tchernyshyov and Chern(2005)Tchernyshyov, and
  Chern]{tchernyshyov2005fractional}
Tchernyshyov,~O.; Chern,~G.-W. Fractional Vortices and Composite Domain Walls
  in Flat Nanomagnets. \emph{Phys. Rev. Lett.} \textbf{2005}, \emph{95},
  197204\relax
\mciteBstWouldAddEndPuncttrue
\mciteSetBstMidEndSepPunct{\mcitedefaultmidpunct}
{\mcitedefaultendpunct}{\mcitedefaultseppunct}\relax
\EndOfBibitem
\bibitem[Pushp \latin{et~al.}(2013)Pushp, Phung, Rettner, Hughes, Yang, Thomas,
  and Parkin]{pushp2013domain}
Pushp,~A.; Phung,~T.; Rettner,~C.; Hughes,~B.~P.; Yang,~S.-H.; Thomas,~L.;
  Parkin,~S. S.~P. Domain Wall Trajectory Determined by Its Fractional
  Topological Edge Defects. \emph{Nat. Phys.} \textbf{2013}, \emph{9},
  505--511\relax
\mciteBstWouldAddEndPuncttrue
\mciteSetBstMidEndSepPunct{\mcitedefaultmidpunct}
{\mcitedefaultendpunct}{\mcitedefaultseppunct}\relax
\EndOfBibitem
\bibitem[Thiaville \latin{et~al.}(2003)Thiaville, Garc{\'\i}a, Miltat, and
  Schrefl]{thiaville2003micromagnetic}
Thiaville,~A.; Garc{\'\i}a,~J.~M.; Miltat,~J.; Schrefl,~T. Micromagnetic Study
  of Bloch-Point-Mediated Vortex Core Reversal. \emph{Phys. Rev. B}
  \textbf{2003}, \emph{67}, 094410\relax
\mciteBstWouldAddEndPuncttrue
\mciteSetBstMidEndSepPunct{\mcitedefaultmidpunct}
{\mcitedefaultendpunct}{\mcitedefaultseppunct}\relax
\EndOfBibitem
\bibitem[Wintz \latin{et~al.}(2016)Wintz, Tiberkevich, Weigand, Raabe, Lindner,
  Erbe, Slavin, and Fassbender]{wintz2016magnetic}
Wintz,~S.; Tiberkevich,~V.; Weigand,~M.; Raabe,~J.; Lindner,~J.; Erbe,~A.;
  Slavin,~A.; Fassbender,~J. Magnetic Vortex Cores as Tunable Spin-Wave
  Emitters. \emph{Nat. Nanotechnol.} \textbf{2016}, \emph{11}, 948--953\relax
\mciteBstWouldAddEndPuncttrue
\mciteSetBstMidEndSepPunct{\mcitedefaultmidpunct}
{\mcitedefaultendpunct}{\mcitedefaultseppunct}\relax
\EndOfBibitem
\bibitem[Ognev \latin{et~al.}(2020)Ognev, Kolesnikov, Kim, Cha, Sadovnikov,
  Nikitov, Soldatov, Talapatra, Mohanty, Mruczkiewicz, Ge, Kerber, Dittrich,
  Virnau, Kl\"{a}ui, Kim, and Samardak]{ognev2020magnetic}
Ognev,~A.~V.; Kolesnikov,~A.~G.; Kim,~Y.~J.; Cha,~I.~H.; Sadovnikov,~A.~V.;
  Nikitov,~S.~A.; Soldatov,~I.~V.; Talapatra,~A.; Mohanty,~J.;
  Mruczkiewicz,~M.; Ge,~Y.; Kerber,~N.; Dittrich,~F.; Virnau,~P.;
  Kl\"{a}ui,~M.; Kim,~Y.~K.; Samardak,~A.~S. Magnetic Direct-Write Skyrmion
  Nanolithography. \emph{ACS Nano} \textbf{2020}, \emph{14}, 14960--14970\relax
\mciteBstWouldAddEndPuncttrue
\mciteSetBstMidEndSepPunct{\mcitedefaultmidpunct}
{\mcitedefaultendpunct}{\mcitedefaultseppunct}\relax
\EndOfBibitem
\bibitem[Zheng \latin{et~al.}(2017)Zheng, Li, Wang, Song, Jin, Wei, Kov{\'a}cs,
  Zang, Tian, Zhang, Du, and Dunin-Borkowski]{zheng2017direct}
Zheng,~F.; Li,~H.; Wang,~S.; Song,~D.; Jin,~C.; Wei,~W.; Kov{\'a}cs,~A.;
  Zang,~J.; Tian,~M.; Zhang,~Y.; Du,~H.; Dunin-Borkowski,~R.~E. Direct Imaging
  of a Zero-Field Target Skyrmion and Its Polarity Switch in a Chiral Magnetic
  Nanodisk. \emph{Phys. Rev. Lett.} \textbf{2017}, \emph{119}, 197205\relax
\mciteBstWouldAddEndPuncttrue
\mciteSetBstMidEndSepPunct{\mcitedefaultmidpunct}
{\mcitedefaultendpunct}{\mcitedefaultseppunct}\relax
\EndOfBibitem
\bibitem[Zeissler \latin{et~al.}(2018)Zeissler, Finizio, Shahbazi, Massey,
  Al~Ma’Mari, Bracher, Kleibert, Rosamond, Linfield, Moore, Raabe, Burnell,
  and Marrows]{zeissler2018discrete}
Zeissler,~K.; Finizio,~S.; Shahbazi,~K.; Massey,~J.; Al~Ma’Mari,~F.;
  Bracher,~D.~M.; Kleibert,~A.; Rosamond,~M.~C.; Linfield,~E.~H.; Moore,~T.~A.;
  Raabe,~J.; Burnell,~G.; Marrows,~C.~H. Discrete Hall Resistivity Contribution
  from N{\'e}el Skyrmions in Multilayer Nanodiscs. \emph{Nat. Nanotechnol.}
  \textbf{2018}, \emph{13}, 1161--1166\relax
\mciteBstWouldAddEndPuncttrue
\mciteSetBstMidEndSepPunct{\mcitedefaultmidpunct}
{\mcitedefaultendpunct}{\mcitedefaultseppunct}\relax
\EndOfBibitem
\bibitem[Guo \latin{et~al.}(2013)Guo, Belova, and
  McMichael]{guo2013spectroscopy}
Guo,~F.; Belova,~L.~M.; McMichael,~R.~D. Spectroscopy and Imaging of Edge Modes
  in Permalloy Nanodisks. \emph{Phys. Rev. Lett.} \textbf{2013}, \emph{110},
  017601\relax
\mciteBstWouldAddEndPuncttrue
\mciteSetBstMidEndSepPunct{\mcitedefaultmidpunct}
{\mcitedefaultendpunct}{\mcitedefaultseppunct}\relax
\EndOfBibitem
\bibitem[Nance \latin{et~al.}(2020)Nance, Roxy, Bhanja, and
  Carman]{nance2020spin}
Nance,~J.~A.; Roxy,~K.~A.; Bhanja,~S.; Carman,~G.~P. Spin-Orbit Torque and
  Dipole Coupling for Nanomagnetic Array Programmability. \emph{IEEE Trans.
  Magn.} \textbf{2020}, \relax
\mciteBstWouldAddEndPunctfalse
\mciteSetBstMidEndSepPunct{\mcitedefaultmidpunct}
{}{\mcitedefaultseppunct}\relax
\EndOfBibitem
\bibitem[Donahue(1999)]{donahue1999oommf}
Donahue,~M.~J. \emph{OOMMF User’s Guide, Version 1.0}; US Department of
  Commerce, National Institute of Standards and Technology, 1999\relax
\mciteBstWouldAddEndPuncttrue
\mciteSetBstMidEndSepPunct{\mcitedefaultmidpunct}
{\mcitedefaultendpunct}{\mcitedefaultseppunct}\relax
\EndOfBibitem
\bibitem[Hartmann(1989)]{hartmann1989point}
Hartmann,~U. The Point Dipole Approximation in Magnetic Force Microscopy.
  \emph{Phys. Lett. A} \textbf{1989}, \emph{137}, 475--478\relax
\mciteBstWouldAddEndPuncttrue
\mciteSetBstMidEndSepPunct{\mcitedefaultmidpunct}
{\mcitedefaultendpunct}{\mcitedefaultseppunct}\relax
\EndOfBibitem
\bibitem[Lohau \latin{et~al.}(1999)Lohau, Kirsch, Carl, Dumpich, and
  Wassermann]{lohau1999quantitative}
Lohau,~J.; Kirsch,~S.; Carl,~A.; Dumpich,~G.; Wassermann,~E.~F. Quantitative
  Determination of Effective Dipole and Monopole Moments of Magnetic Force
  Microscopy Tips. \emph{J. Appl. Phys.} \textbf{1999}, \emph{86},
  3410--3417\relax
\mciteBstWouldAddEndPuncttrue
\mciteSetBstMidEndSepPunct{\mcitedefaultmidpunct}
{\mcitedefaultendpunct}{\mcitedefaultseppunct}\relax
\EndOfBibitem
\bibitem[Magiera and Schulz(2014)Magiera, and Schulz]{magiera2014magnetic}
Magiera,~M.~P.; Schulz,~S. Magnetic Vortices Induced by a Monopole Tip.
  \emph{IEEE Trans. Magn.} \textbf{2014}, \emph{50}, 1--4\relax
\mciteBstWouldAddEndPuncttrue
\mciteSetBstMidEndSepPunct{\mcitedefaultmidpunct}
{\mcitedefaultendpunct}{\mcitedefaultseppunct}\relax
\EndOfBibitem
\bibitem[Vansteenkiste \latin{et~al.}(2014)Vansteenkiste, Leliaert, Dvornik,
  Helsen, Garcia-Sanchez, and Van~Waeyenberge]{vansteenkiste2014design}
Vansteenkiste,~A.; Leliaert,~J.; Dvornik,~M.; Helsen,~M.; Garcia-Sanchez,~F.;
  Van~Waeyenberge,~B. The Design and Verification of MuMax3. \emph{AIP Adv.}
  \textbf{2014}, \emph{4}, 107133\relax
\mciteBstWouldAddEndPuncttrue
\mciteSetBstMidEndSepPunct{\mcitedefaultmidpunct}
{\mcitedefaultendpunct}{\mcitedefaultseppunct}\relax
\EndOfBibitem
\bibitem[Venkat \latin{et~al.}(2018)Venkat, Fangohr, and
  Prabhakar]{venkat2018absorbing}
Venkat,~G.; Fangohr,~H.; Prabhakar,~A. Absorbing Boundary Layers for Spin Wave
  Micromagnetics. \emph{J. Magn. Magn. Mater.} \textbf{2018}, \emph{450},
  34--39\relax
\mciteBstWouldAddEndPuncttrue
\mciteSetBstMidEndSepPunct{\mcitedefaultmidpunct}
{\mcitedefaultendpunct}{\mcitedefaultseppunct}\relax
\EndOfBibitem
\bibitem[hpc()]{hpc}
Research Computing Service.
  \url{https://www.imperial.ac.uk/admin-services/ict/self-service/research-support/rcs/}
  (accessed Dec 2, 2020), DOI: 10.14469/hpc/2232\relax
\mciteBstWouldAddEndPuncttrue
\mciteSetBstMidEndSepPunct{\mcitedefaultmidpunct}
{\mcitedefaultendpunct}{\mcitedefaultseppunct}\relax
\EndOfBibitem
\bibitem[Kl{\"a}ui \latin{et~al.}(2003)Kl{\"a}ui, Vaz, Rothman, Bland,
  Wernsdorfer, Faini, and Cambril]{klaui2003domain}
Kl{\"a}ui,~M.; Vaz,~C. A.~F.; Rothman,~J.; Bland,~J. A.~C.; Wernsdorfer,~W.;
  Faini,~G.; Cambril,~E. Domain Wall Pinning in Narrow Ferromagnetic Ring
  Structures Probed by Magnetoresistance Measurements. \emph{Phys. Rev. Lett.}
  \textbf{2003}, \emph{90}, 097202\relax
\mciteBstWouldAddEndPuncttrue
\mciteSetBstMidEndSepPunct{\mcitedefaultmidpunct}
{\mcitedefaultendpunct}{\mcitedefaultseppunct}\relax
\EndOfBibitem
\bibitem[Bogart \latin{et~al.}(2009)Bogart, Atkinson, O’Shea, McGrouther, and
  McVitie]{bogart2009dependence}
Bogart,~L.~K.; Atkinson,~D.; O’Shea,~K.; McGrouther,~D.; McVitie,~S.
  Dependence of Domain Wall Pinning Potential Landscapes on Domain Wall
  Chirality and Pinning Site Geometry in Planar Nanowires. \emph{Phys. Rev. B}
  \textbf{2009}, \emph{79}, 054414\relax
\mciteBstWouldAddEndPuncttrue
\mciteSetBstMidEndSepPunct{\mcitedefaultmidpunct}
{\mcitedefaultendpunct}{\mcitedefaultseppunct}\relax
\EndOfBibitem
\end{mcitethebibliography}
\begin{suppinfo}

\subsection*{Animations of state-control protocols}
Contained within this folder are animations of the state-control protocols for all combinations of starting magnetisation state and tip polarity. In each case, nanodisk dimensions of 150 nm width $\times$ 10 nm thickness with the exception of `Regime\_E\_pol\_switch.avi' which has nanodisk dimensions of 250 nm width $\times$ 20 nm thickness. The title of the videos have the following format - `Starting state Final state Tip polarity'.

\subsection*{State-control in terms of topological defects}
More insight into the micromagnetics of the state-control protocols can be achieved through a description of the evolution of topological defects\cite{pushp2013domain,tchernyshyov2005fractional} - locations at which spins diverge from a collinear state. The texture of this divergence is described by a winding number \textit{w\textsubscript{n}}. For edge-bound topological defects, \textit{w\textsubscript{n}} is fractional whereas for topological defects free to move in the bulk (\textit{e.g.} vortices), \textit{w\textsubscript{n}} is an integer [Fig. 2 b-d)]. In ferromagnetic spin systems, \textit{w\textsubscript{n}} is conserved and in a nanodisk, must sum to 1.

Nanodisks in a macrospin state possess two +$\frac{1}{2}$ \textit{w\textsubscript{n}} defects shown in Fig. 2 e-g). As H\textsubscript{tip} traverses the surface of the disk the local spins align with the radial-field profile forming a vortex-like texture (+1 \textit{w\textsubscript{n}}) underneath the tip leaving a -$\frac{1}{2}$ edge-bound defect in its wake [Fig. 2 e), t = 0.65 ns]. The +1 vortex defect underneath the tip and the edge-bound -$\frac{1}{2}$ defect are bound by a chain of reversed spins which increases in size and energy as the tip continues across the surface. To reduce this energy, the -$\frac{1}{2}$ defect traverses the edge of the nanodisk along the lowest energy pathway determined by the offset of the tip trajectory relative to the axis of movement [Fig. 2 f) t = 1 ns, g) t = 1.1 ns]. Relaxation of spins away from the tip facilitate the movement of the +$\frac{1}{2}$ defect towards the chain of reversed spins. Eventually, the $\pm \frac{1}{2}$ edge-bound defects annihilate leaving the nanodisk in a vortex state where the chirality is determined by the direction of defect movement [Fig. 2 f) t = 1.1 ns, g) t = 1.2 ns]. As the tip moves towards the edge of the disk, energy minimisation results in the formation of another $\pm \frac{1}{2}$ defect pair [Fig. 2 e) t = 1.6 ns]. As the tip moves away from the edge of the disk the +1 defect can no longer follow. Instead the +1 and -$\frac{1}{2}$ defect combine to form a +$\frac{1}{2}$ defect at the edge of the disk. This defect then traverses around the disk-edge resulting in a macrospin state. 

\begin{figure}[h!]
\centering

\includegraphics[width=\textwidth]{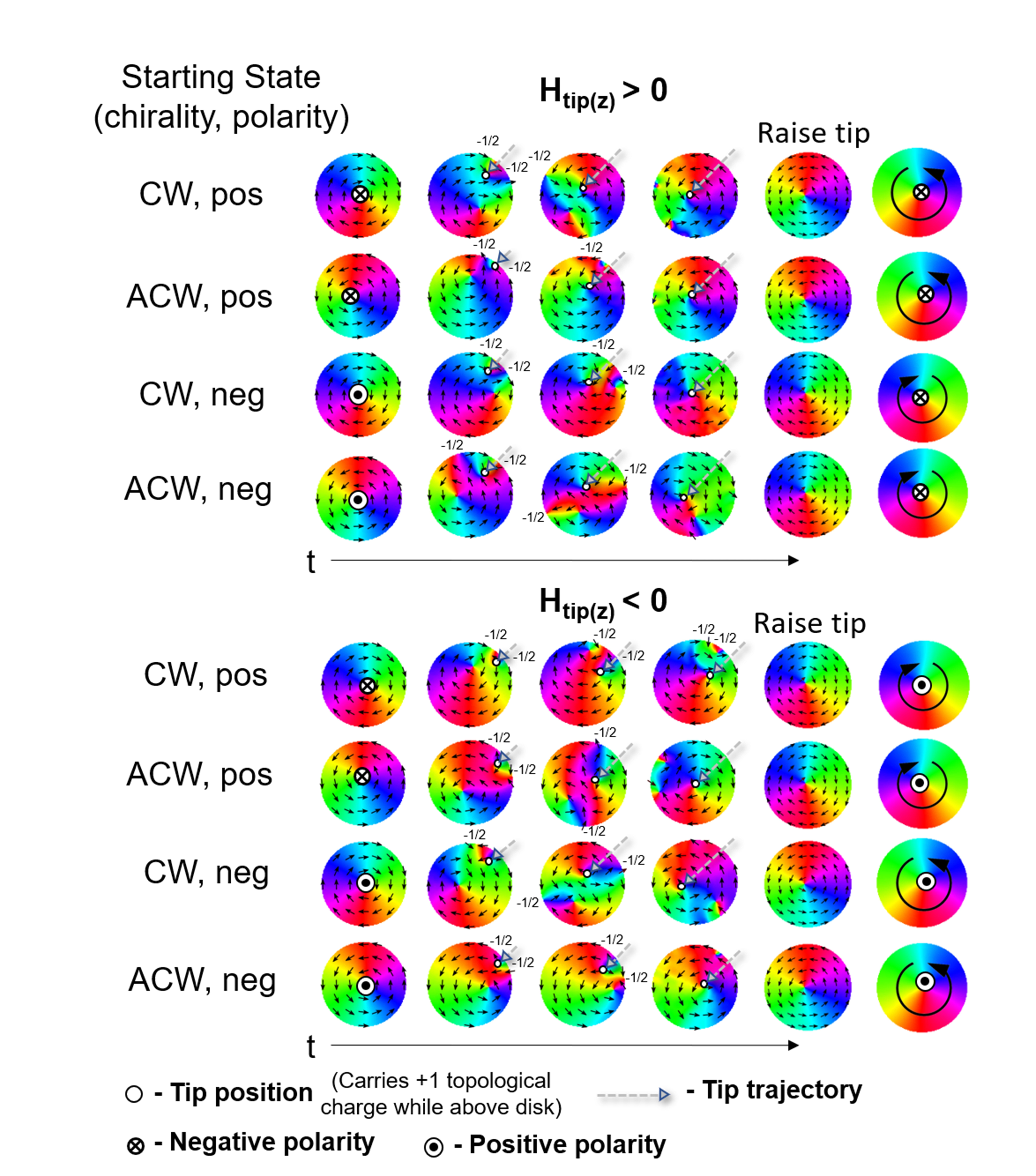}
\caption{Time evolution series of the state-control protocols with all combinations of vortex starting state and tip polarity. The dynamics during state-control and final state depend on the initial microstate and tip polarity. When the polarity of the tip and the vortex core are aligned, the resulting chirality is anti-clockwise. If they are anti-aligned, the chirality is clockwise. For all cases, the final polarity aligns with the polarity of the tip. Disk dimensions of width = 150 nm and thickness = 10 nm were used.}
\label{Fig.vortex} \vspace{-1em}
\end{figure}
For vortex starting states [Fig. 2 h), Fig \ref{Fig.vortex}], we start with a +1 topological defect which is free to move in the bulk. The addition of another tip bound +1 vortex is accompanied by the formation of two -$\frac{1}{2}$ edge-bound defects. Depending on the polarity of the tip and initial chirality and polarity of the vortex, the original +1 defect is either attracted-to or repelled-by the tip. If the vortex is repelled by the tip, a chirality reversal occurs whereas if the vortex core is attracted to the tip, no chirality switching occurs. As time progresses, the two -$\frac{1}{2}$ defects traverse the edge of the disk until they annihilate with the initial +1 vortex leaving a tip-bound vortex state where the vortex core polarity is determined by the tip polarity. This leaves the nanodisk in a vortex state from which the process is the same as in the macrospin starting case. 

\begin{figure}[htb]
\centering

\includegraphics[width=\textwidth]{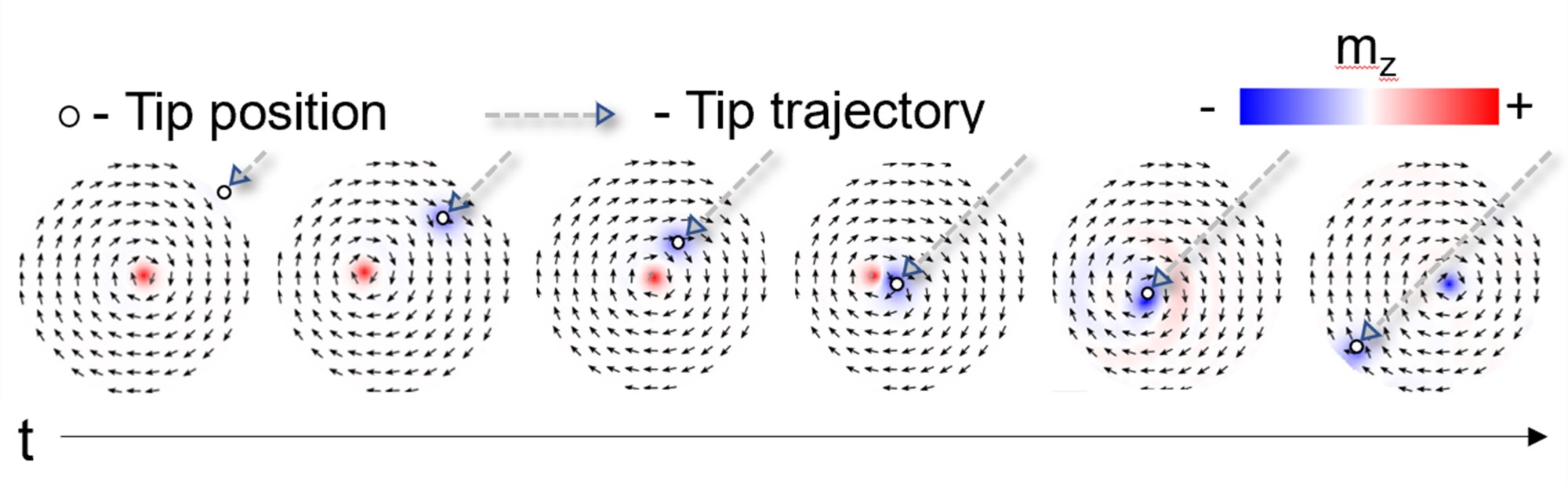}
\caption{Time evolution series of vortex polarity switching in region \textbf{E} of Fig 2 b). In this regime, a vortex is no longer induced under the tip. Instead, polarity switching occurs as a result of minimising the Zeeman energy when the tip is in the vicinity of the vortex core. Disk dimensions of width = 250 nm and thickness = 20 nm were used.}
\label{Fig.E} \vspace{-1em}
\end{figure}
For larger disk dimensions [Fig. 2  b) Regime \textbf{E}, Fig. \ref{Fig.E}], as the tip traverses the surface of the nanodisk, a vortex no longer forms, instead there is only a slight distortion in the spin texture. If the vortex core polarity and the tip polarity are of opposite sign and H\textsubscript{tip} has sufficient strength, the vortex core polarity reverses to align with H\textsubscript{tip} minimising the Zeeman energy in the process. The upper limit of this region is determined by the thickness of the nanodisk and the magnitude of H\textsubscript{tip}.

\subsection*{State-control in non-ideal disks}
The results previously presented have focused on state-control in ideal disks. However, fabrication methods inherently produce non-idealities such as asymmetry and edge roughness. As such, it is important to consider these effects and how they may affect the state-control method. Provided that the final state is magnetically stable in the disk (whether perfectly circular or possessing some degree of asymmetry), all the writing operations we describe will remain possible due to the conservation of net winding number in the disk. The vortex chirality depends only on whether the +1/2 charge approaches the -1/2 charge in a clockwise or anti-clockwise fashion, hence the trajectories of the half-integer defects are what determines the final state, rather than the final state energy relative to that of the alternative states. As such, any asymmetry or roughness will then affect the motion of the half-defects which ultimately determines the chirality of the resulting vortex.  
Fig. \ref{Fig.ellipse} shows multiple time-evolution plots of writing protocols performed on an ellipse with a width and length of 150 nm and 180 nm respectively. When scanning parallel to the long axis of the ellipse [Fig. \ref{Fig.ellipse} a-c)], the system remains symmetric along the tip trajectory and the final state is prepared as in a circular nanodisk. When approaching from 45$^{\circ}$ to the long axis [Fig. \ref{Fig.ellipse} d-g)], there is now a preferred direction for half-defect rotation, and thus the system favours a given chirality (in this case anti-clockwise), with a larger region of the final tip position phase space resulting in an ACW final state. However, even in this case of a highly asymmetric nanoelement, a substantial region of this phase space remains which results in a CW final state, hence the less favourable chirality state may still be prepared by choosing a suitable final tip position. [Fig. \ref{Fig.ellipse} f,g)]. When scanning along the short axis of the ellipse [Fig. \ref{Fig.ellipse} h-k)], the initial macrospin magnetisation no longer aligns with the radial H\textsubscript{tip} profile. This again results in one chirality being favoured. The other chirality may be achieved by suitably offsetting the tip or by scanning in the opposite direction [Fig. \ref{Fig.ellipse} h)]. In all cases, the final macrospin axis always realigns with H\textsubscript{tip} and the vortex polarity aligns with the tip polarity. It is important to note that the chosen ellipse dimensions are an extreme case and any asymmetry when fabricating circular-shaped nanoelements is expected to be far less.

Asymmetry may also take the form of a D-shaped disk with a flattened edge. This nanoelement shape has been shown to favour one sense of chirality when applying a global field \cite{dumas2011chirality}. Fig. \ref{Fig.d_shaped} shows the state-control protocol performed on a D-shaped disk with a diameter of 150 nm, thickness of 10 nm and an edge flattened at a point 30 nm short of the edge of a 150 nm diameter circle. Again, by offsetting the tip relative to the scan direction, both vortex chiralities can be achieved. These results emphasise the importance of the topological nature of the state-control protocols. 

With regards to edge roughness, pinning sites may affect the complete rotational freedom of the  macrospin state as local energy minima may result in the macrospin relaxing slightly away from the desired axis which may result in a preferential direction for half-integer defect movement. As previously, suitably offsetting the tip will allow both chiralities to be prepared. This will be further reduced in dense arrays as the strong dipolar coupling between neighbouring nanodisks in a macrospin state will favour the alignment of neighbouring macrospin magnetisation. Pinning sites introduced from fabrication and material defects introduce some aspect of irregularity to the precise dynamics of the writing process, but the topologically protected nature of the writing process is expected to be enough to result in the same final states in all but the most extreme of cases, as evidenced by the experimentally verified efficacy of the related tip writing processes demonstrated in previous publications \cite{gartside2016novel,gartside2018realization}. Furthermore, edge roughness in the form of notches may also affect the final state by pinning certain half-defects and preventing them from traversing the edge of the nanodisk and annihilating. A full investigation of this is beyond the scope of this work, however notch-induced domain wall pinning in nanowires typically requires a notch size much larger in depth than is expected from fabrication induced defects \cite{klaui2003domain,bogart2009dependence}.

\begin{figure}[h!]
\centering
\includegraphics[width=\textwidth]{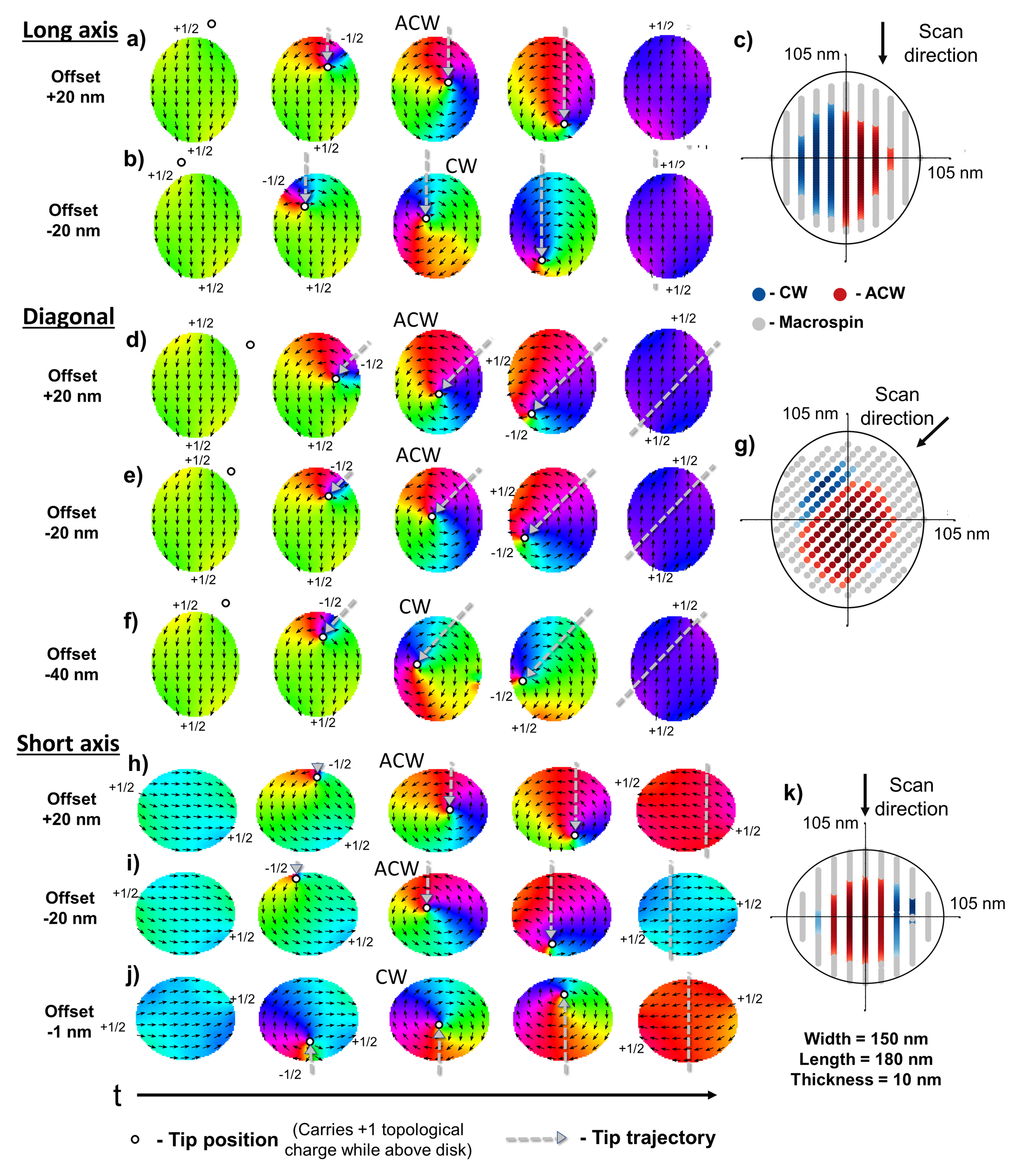}
\caption{Time evolution series of state-control on an elliptical nanoelement with short axis, long axis and thickness dimensions of 150 nm, 180 nm and 10 nm respectively. a-c)When scanning parallel to the long axis of the ellipse, the system remains symmetric along the tip trajectory and the final state is prepared as in a circular nanodisk. When approaching from  d-g) 45 degrees to the long axis or h-j) perpendicular to the long axis, there is now a preferred direction for half-defect rotation, and thus the system favours a given chirality (in this case ACW). In each case, all combinations of vortex polarity and chirality can be achieved by modifying the scan offset and direction.}
\label{Fig.ellipse} \vspace{-1em}
\end{figure}
\begin{figure}[h!]
\centering
\includegraphics[width=\textwidth]{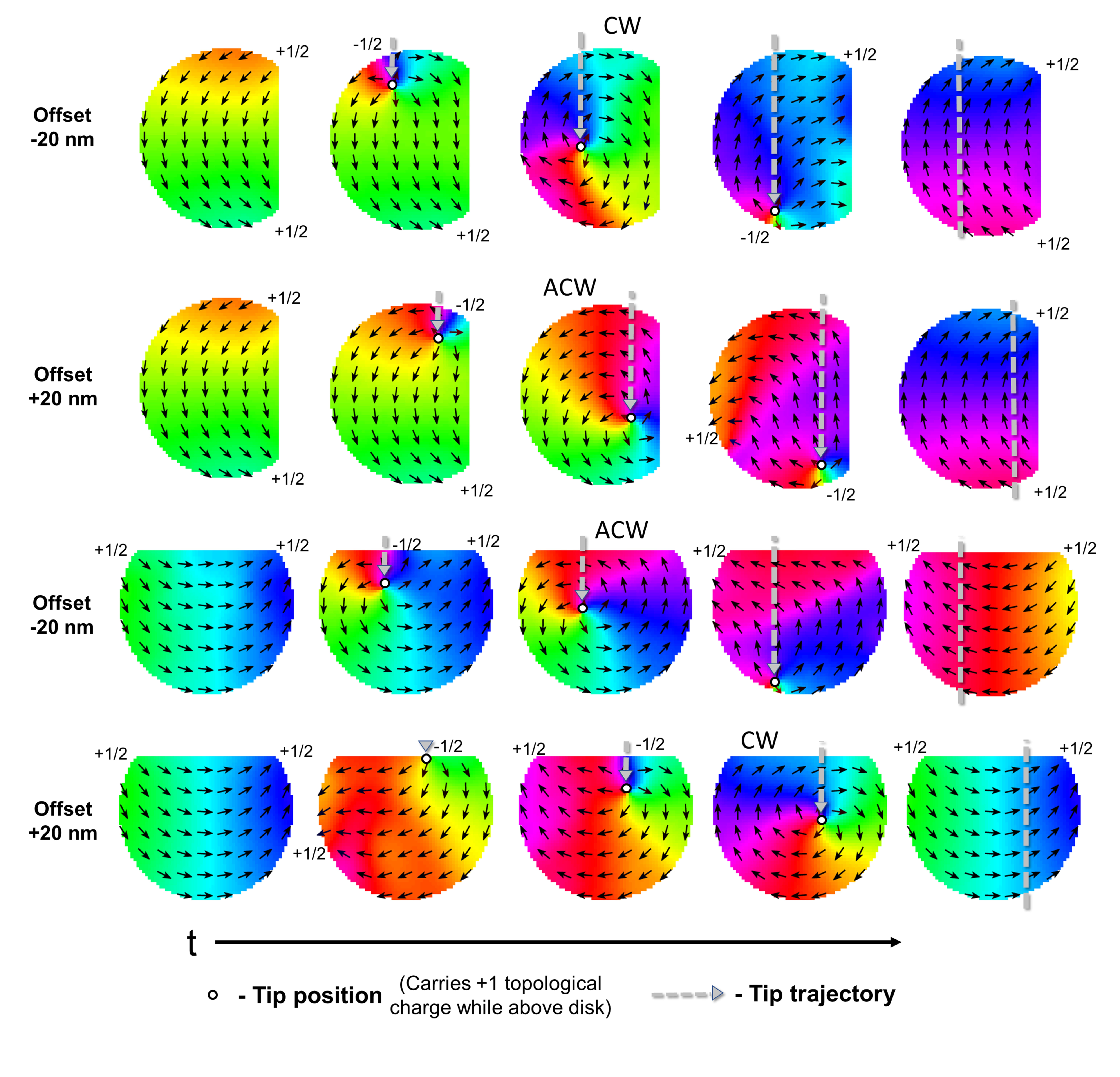}
\caption{Time evolution series of state-control on a d-shaped nanodisk with a diameter of 150 nm, thickness of 10 nm and a flattened right edge (30 nm). All combinations of vortex chirality and polarity can be achieved.}
\label{Fig.d_shaped} \vspace{-1em}
\end{figure}
\subsection*{State-control in nanodisk arrays}
In the previous results, the tip approaches a nanodisk from a distance of 400 nm in the xy-plane and subsequently performs a given operation. However, in dense arrays, approaching from this distance may not be feasible without affecting neighbouring nanodisks. Instead, the tip may be lowered next to the disk edge before performing a given protocol. In Fig. \ref{Fig.above_sinlge}, the tip is approached from a height of 1 $\mu$m to 15 nm from the top right edge of the nanodisk and subsequently scanned across the nanodisk surface. For both macrospin and vortex starting states, the dynamics of the state-control protocols are equivalent to the xy-approach. In Fig. \ref{Fig.above_array}, this method is extended to a 5 x 5 array of nanodisks with a disk diameter of 150 nm and an inter-disk separation of 25 nm each starting in a vortex state. The tip is first lowered to the top right edge of the central nanodisk. A macrospin state is then written by scanning across the disk surface and raising tip beyond the bottom-left edge. Next, the tip is lowered to the bottom-right edge during which the macrospin aligns with H\textsubscript{tip}. The tip is then scanned and raised above the centre of the central nanodisk leaving a vortex state. This demonstrates that the state-control protocols can be used to control individual nanodisk states in an array without disturbing the surrounding magnetic states.

\begin{figure}[h!]
\centering

\includegraphics[width=\textwidth]{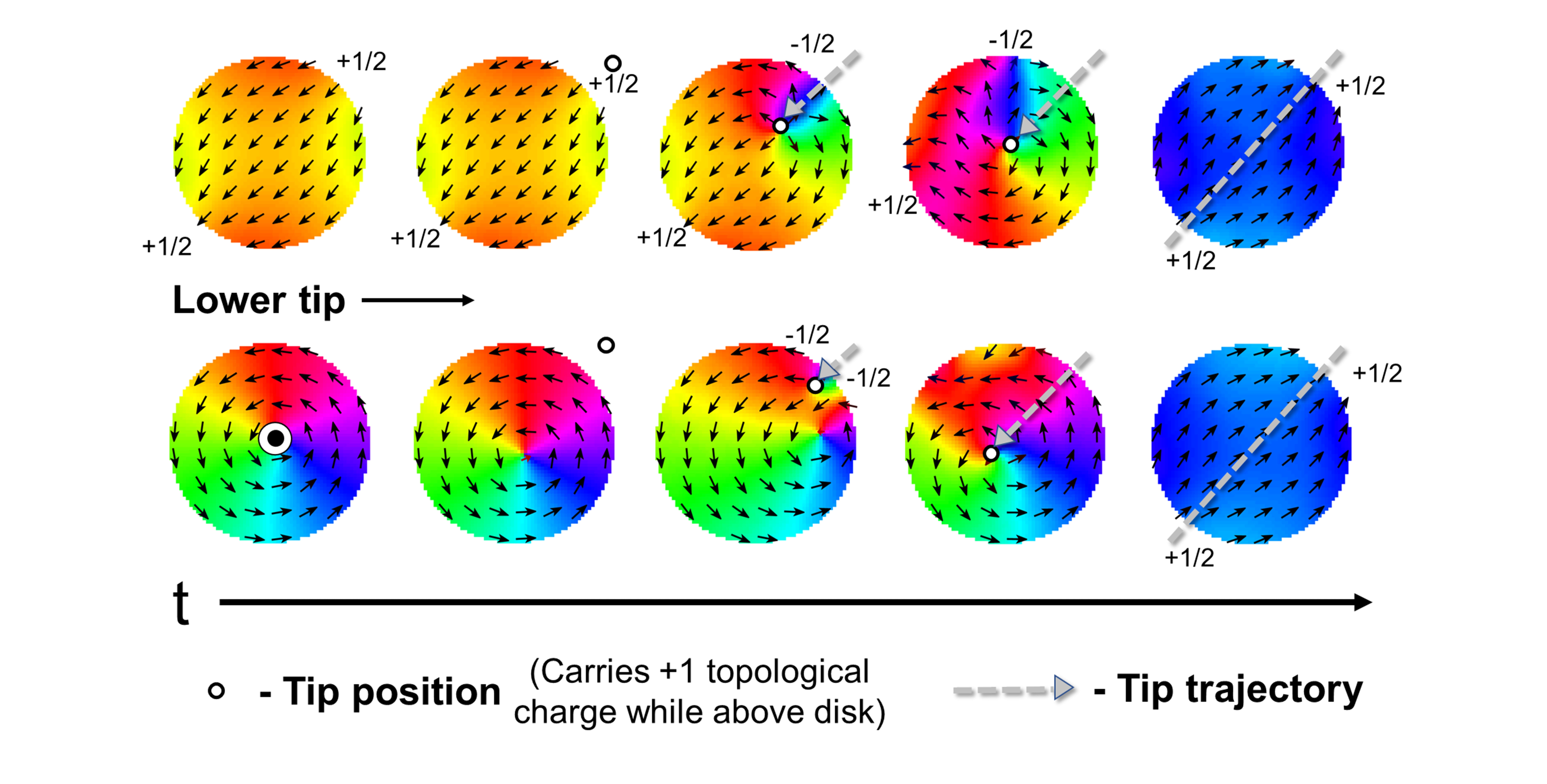}
\caption{Time evolution series of preparing nanodisk states where the tip is lowered from 1 um to the top-right edge of the nanodisk and subsequently scanned across the nanodisk surface. The dynamics and final state of the protocol is equivalent to when the tip approaches in the xy-plane. The disk diameter and thickness are 150 nm and 10 nm respectively.}
\label{Fig.above_sinlge} \vspace{-1em}
\end{figure}

\begin{figure}[h!]
\centering

\includegraphics[width=\textwidth]{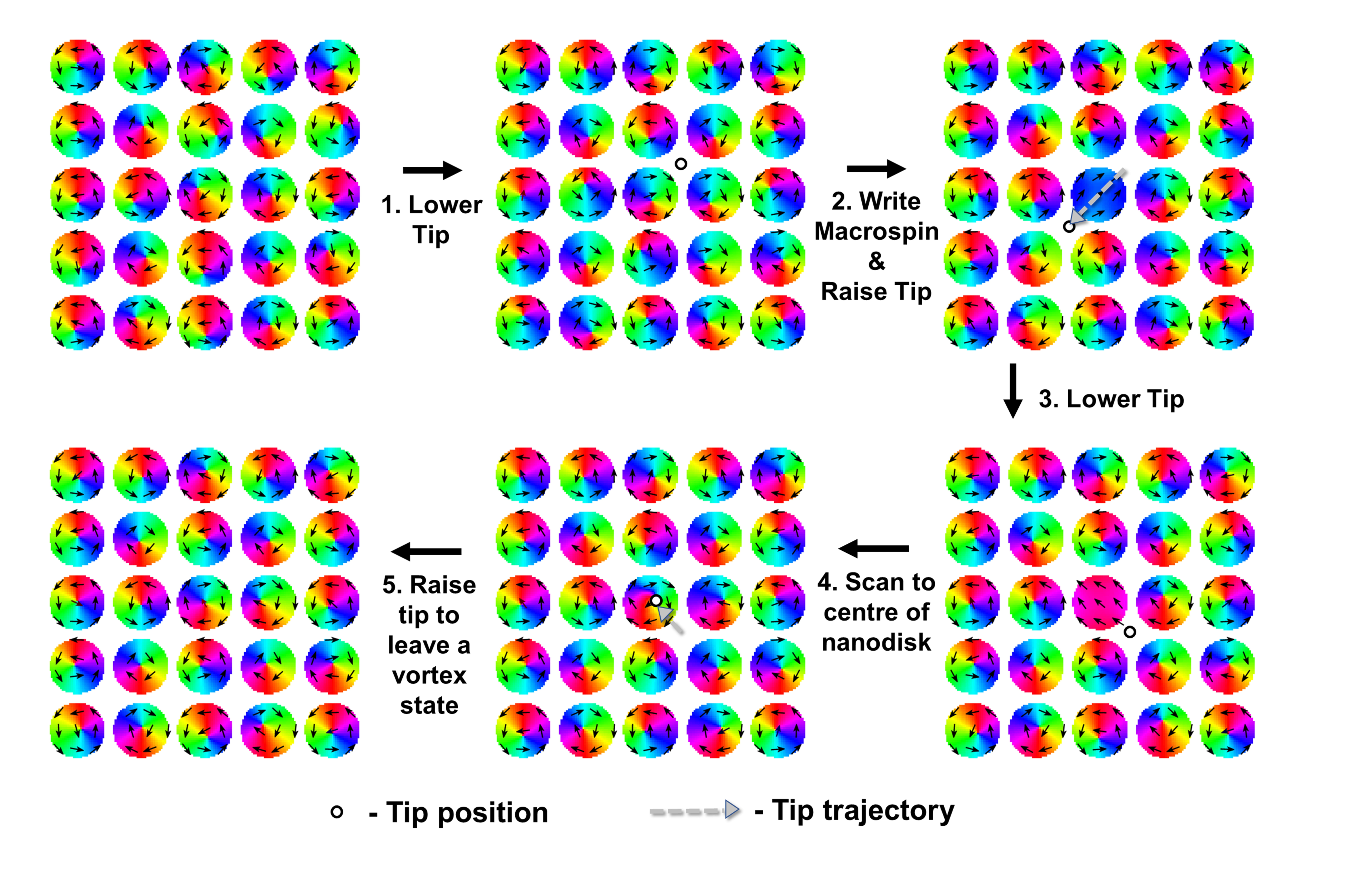}
\caption{A demonstration of state-control in a dense nanodisk array in a with disk diameter, thickness and inter-disk separation of 150 nm, 10 nm and 25 nm respectively with each disk initialised in a vortex state. 1. The tip is first lowered from 1 $\mu$m to the top right edge of the central disk. 2. The tip is then scanned across the nanodisk to write a macrospin state. 3. The tip is then lowered to the bottom right edge, during which the macrospin realigns with H\textsubscript{tip}. 4. The tip is then scanned across to the centre of the central nanodisk. 5. The tip is raised above the centre of the nanodisk leaving a vortex state.}
\label{Fig.above_array} \vspace{-1em}
\end{figure}

\subsection*{Comparison of excitation methods}
In this study, two excitation methods were used to probe the dynamics of nanodisk-based RMCs. For the waveguiding and gating presented in Fig. 5, a sinc pulse exciting modes up to 25 GHz was applied to the first four columns of nanodisks whereas for the remaining simulations, a continuous \textit{f} = 2.3 GHz sinusoidal field was applied to the first two nanodisks. Fig. \ref{Fig-excitation} shows the spin-wave spectra averaged over each disk along a straight macrospin pathway. For the sinc-pulse excitation, a broad band of frequencies are initially excited between 1 - 3.5 GHz. These frequencies then converge as the spin-waves propagate along the pathway. The sinusoidal field initially excites a smaller band of frequencies around the edge mode resonance. We also see the excitation of higher frequencies at multiples of the excitation frequency. Both methods provide similar frequency spectra along the macrospin pathway.
\begin{figure}[h!]
\centering

\includegraphics[width=\textwidth]{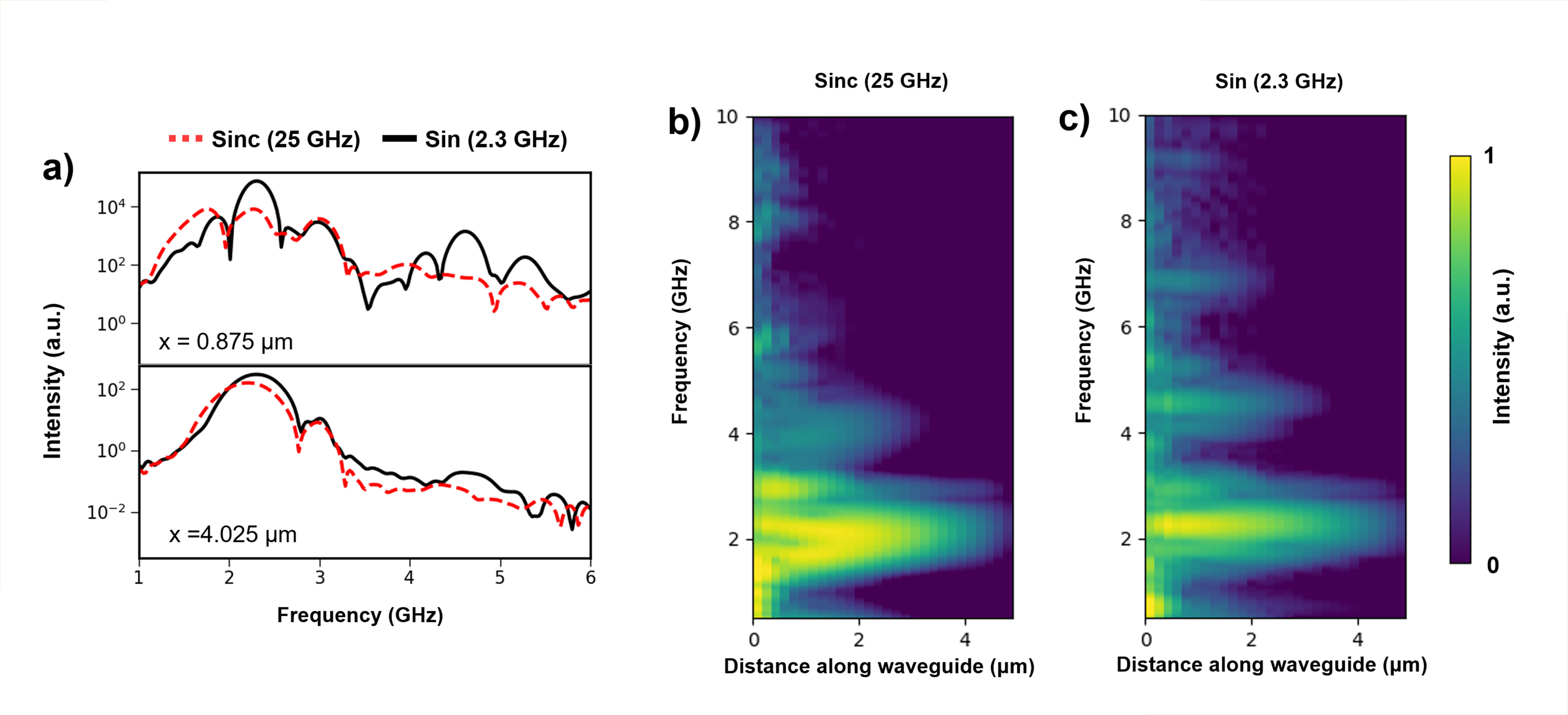}
\caption{a) Comparison of the spin-wave spectra for the two excitation methods at a distance of 0.875 $\mu$m (top panel) and 4.025 $\mu$m (bottom panel) along the waveguide. Spin wave spectra along a straight waveguide for b) a sinc-pulse exciting the first four nanodisks and c) a continuous \textit{f} = 2.3 GHz sinusoidal field on the first two nanodisks used in this work.}
\label{Fig-excitation} \vspace{-1em}
\end{figure}

\subsection*{Magnon splitting}

In Fig. 6 it was shown that there is a preferential direction for power transmission when the macrospin pathway is subjected to a 180$^{\circ}$ split. Equal splitting is achieved when there are two or more additional macrospin nanodisks placed along the x-direction. Fig. \ref{Fig-cross} shows that when one additional macrospin disk is added, the junction remains asymmetric resulting in a preferential power splitting with a difference of 9 dB between the 'up' and 'down' branch. 
\begin{figure}[h!]
\centering

\includegraphics[width=\textwidth]{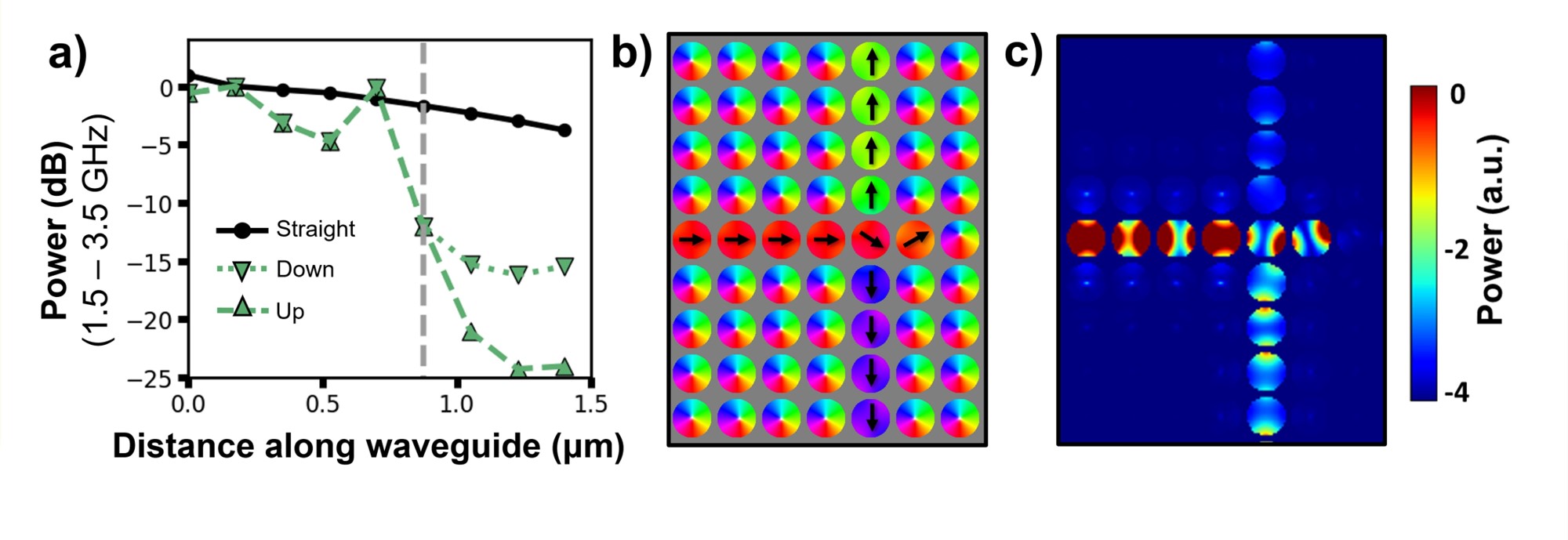}
\caption{a) Power profile, b) Microstate configuration and c) power spectra of magnon splitting across a 180$^{\circ}$ split with the addition of one macrospin disk. In this case, the junction remains asymmetric resulting in a preferential direction for power propagation.}
\label{Fig-cross} \vspace{-1em}
\end{figure}

\subsection*{Design considerations for logic functionality}
When designing more complex microstate architectures for logic operations such as the interferometer design in Fig. 8, both power losses and phase shifts must be considered in both interferometer arms. The first design step is to introduce an input which causes a $\pi$ phase shift relative to the other input in order to maximise destructive interference in the 01 / 10 cases. The phase shift functionalities presented in Fig. 7 are all accompanied by an additional power losses. As such, the two arms of the interferometer will meet at different powers in the 01 / 10 cases. Considering an interferometer design similar to  Fig. 8 but rather than a power sink as an input of 0, we instead had a straight macrospin pathway with no adjacent macrospin nanodisk, the arm with a state of 0 will always dominate when recombining. Therefore, the 00 state will have a high power, the 01 / 10 will have an intermediate power and the 11 case will have a low power. Therefore, it is necessary to have a second input that attenuates the power to the same level as the phase shift input but does not induce a phase shift itself. For this, we can use the power sink presented in Fig. 6 g). It is important that the power loss induced by both the 0 and 1 states match both arms meet with the same power and any changes in power between the combination of logic states only results from the constructive / destructive interference due to the phase difference.

A different option is to use a power source rather than a power sink to compensate for the power reduction caused by the phase shift. For the design in Fig. 8, the power source would be an additional macrospin pathway feeding into the original pathways which would be excited from one end. It would need to be located on the phase shift branch for each configuration which would require additional state changes, multiple excitation points and must be excited with the optimum power, significantly increasing the overall complexity compared to the current design. 
To reduce the number of state changes, the power source microstates could be permanent with the requirement that a power source is only excited if a phase shift is present. If a phase-shift is not present, then the power source would not be excited. However, this would result in spin waves splitting and propagating down the power source rather than too the intended output. To summarise, careful consideration of the various processes occurring across the entire microstate is necessary to achieve the desired functionality.

\subsection*{Verification of RMC simulation times}
The ‘straight’, ‘gate’ and ‘stop’ simulations presented in Fig. 5 were ran for 12 ns. The bending, splitting and phase shift simulations presented in Fig. 6 and Fig. 7 ran for 8 ns and the interferometry simulations in Fig. 8 ran for 12 ns. In the interferometry simulations, a 2ns slice was taken from the end of the simulation to avoid any initial transients.

For the ‘straight’, ’gate’ and ‘stop’ simulations, the first four columns of nanodisks are excited with an out-of-plane sinc-pulse with a cutoff frequency of 25 GHz. Therefore, a steady state is reached once the magnetisation has relaxed back down to static configuration. Fig. \ref{verification} a) shows the change in magnetisation from the initial static state as a function of simulation time for the ‘straight’ case. At 12 ns, there is a minimal net change in magnetisation across the array thus that the system has relaxed back to its static configuration and an equilibrium has been reached. Fig. \ref{verification} b) shows the output power the final disk in the 'bend' microstate configuration as a function of simulation time. The power presented at 8 ns is similar to those at longer simulation times, hence a steady state has been reached.

Fig. \ref{verification} c) shows m\textsubscript{z} of the output disk as a function of simulation time. Whilst amplitude modulations are present over longer timescales, there remains a clear difference between the 00/11 \& 10/01 states. Fig. \ref{verification} d) shows the output power of the final disk as a function of simulation time. Over the course of the simulation, the states remain clearly distinguishable. 
\begin{figure}[h!]
\centering

\includegraphics[width=\textwidth]{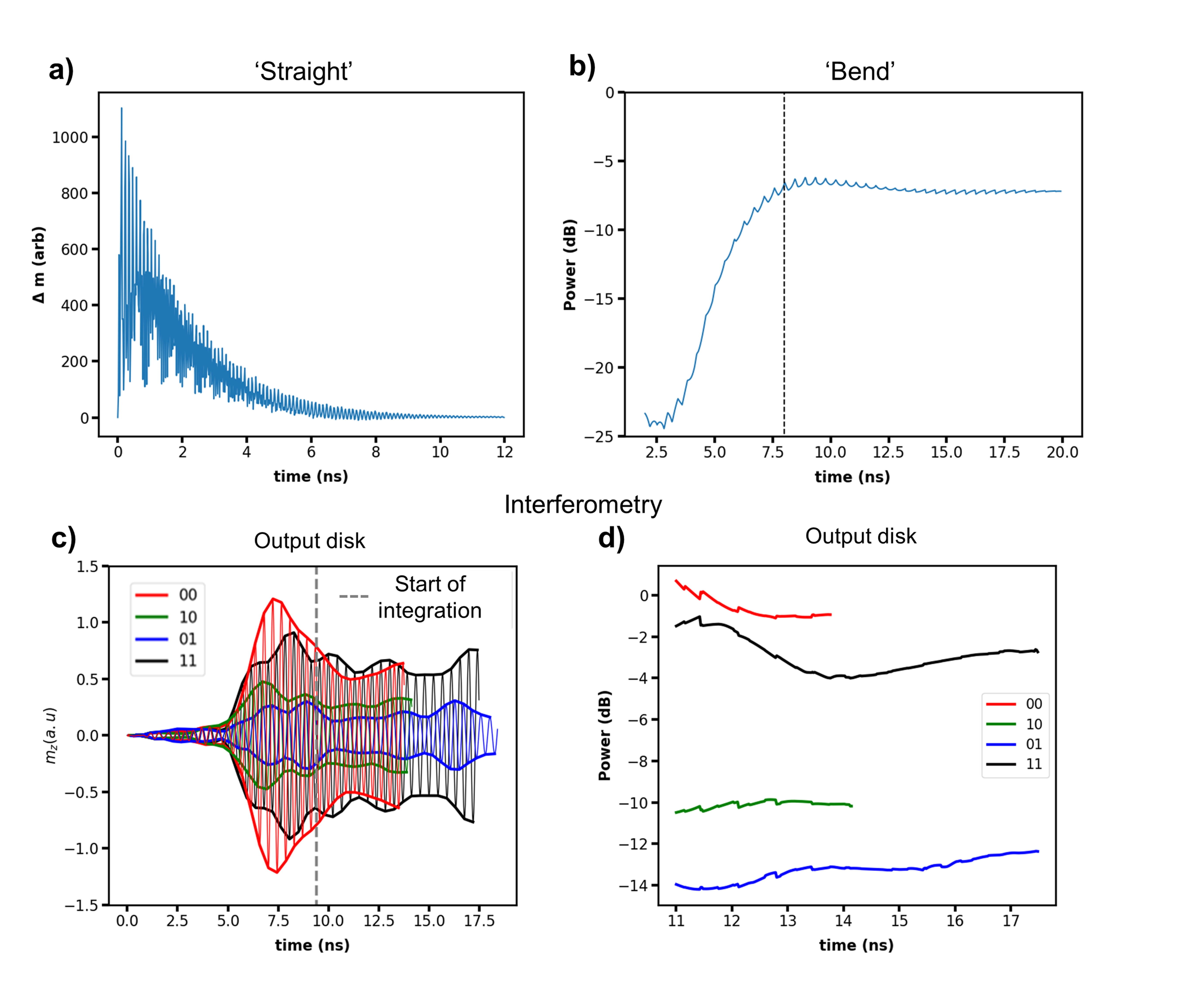}
\caption{Verification of simulation lengths. a) Change in magnetisation over time when exciting a straight macrospin pathway with a sinc-pulse. The magnetisation relaxes after 12 ns. b) The power of the final disk in a ‘bend’ as a function of simulation time. The output at 8 ns is compatible to those at longer timescales. c) Profiles of m\textsubscript{z} vs t for the four interferometry microstates. Whilst there is some modulation, the differences between the states remain clear for longer simulation times. d) Output power as a function of simulation time for the four interferometry microstates obtained by integrating the Fourier transform of the time-domain signals in c). The first 9.7 ns is removed from the integration to avoid the initial transient. The output powers remain differentiable.}
\label{verification} \vspace{-1em}
\end{figure}
\end{suppinfo}

\end{document}